\documentclass{article}

\usepackage{PRIMEarxiv}

\usepackage[utf8]{inputenc} 
\usepackage[T1]{fontenc}    
\usepackage{hyperref}       
\usepackage{url}            
\usepackage{booktabs}       
\usepackage{amsfonts}       
\usepackage{nicefrac}       
\usepackage{microtype}      
\usepackage{lipsum}
\usepackage{fancyhdr}       
\usepackage{graphicx}       %
\usepackage{amsmath}
\usepackage{comment}
\usepackage{tikz}
\usetikzlibrary{positioning, arrows.meta}
\usepackage[utf8]{inputenc} 
\usepackage[T1]{fontenc}    
\usepackage{hyperref}       
\usepackage{url}            
\usepackage{booktabs}       
\usepackage{amsfonts}       
\usepackage{nicefrac}       
\usepackage{microtype}      
\usepackage{xcolor}         
\usepackage{placeins}
\usepackage{subcaption}

\providecommand{\keywords}[1]{\textbf{\textit{Keywords---}} #1}
\newcommand{\scref}[1]{\S\ref{#1}}

\makeatletter
\renewcommand{\section}{%
  \@startsection{section}{1}{\z@}%
  {0.1ex \@plus 0.1ex \@minus 0.1ex}%
  {0.2ex \@plus 0.1ex}%
  {\normalfont\Large\bfseries}%
}

\renewcommand{\subsection}{%
  \@startsection{subsection}{2}{\z@}%
  {0.25ex \@plus 0.4ex \@minus 0.2ex}%
  {0.2ex \@plus 0.1ex}%
  {\normalfont\large\bfseries}%
}

\renewcommand{\subsubsection}{%
  \@startsection{subsubsection}{3}{\z@}%
  {0.1ex \@plus 0.1ex \@minus 0.1ex}
  {0.1ex \@plus 0.1ex}
  {\normalfont\normalsize\bfseries}
}

\renewcommand{\paragraph}{%
  \@startsection{paragraph}{4}%
  {\z@}{0.1ex \@plus 0.05ex \@minus 0.05ex}{-1.5em}%
  {\normalfont\normalsize\bfseries}%
}
\makeatother

\usepackage{enumitem}
\setlist[itemize]{topsep=0pt, partopsep=0pt, itemsep=0pt, parsep=0pt}

\usepackage{longtable,array,calc}
\usepackage{etoolbox}
\makeatletter
\patchcmd\longtable{\par}{\if@noskipsec\mbox{}\fi\par}{}{}
\makeatother
\providecommand{\tightlist}{%
  \setlength{\itemsep}{0pt}\setlength{\parskip}{0pt}}
\title{From Embedding Geometry to Spectral Search: \\Energy Dispersion Networks for Vector Retrieval}

\author{
Lorenzo Moriondo \\
tuned.org.uk - Genefold AI \\
\texttt{tunedconsulting@gmail.com} \\
\texttt{lorenzo@genefold.ai} 
\And
Ilias Azizi \\
LIPADE, Université Paris Cité \\
France\\
\texttt{ilias.azizi@u-pariscite.fr}
}

\begin{document}
\maketitle

\begin{abstract}
High-dimensional vector spaces, particularly embedding spaces with dense semantic structure, are often interpreted primarily leveraging solely geometric relationships. In this work, we show that they can also be viewed as spectral energy networks induced by the topology of their underlying feature-space manifold with relevant improvements for downstream tasks. Building on this perspective, we introduce \textit{Graph Wiring}, a general framework for exploiting feature-space spectral structure, together with \textit{Spectral Indexing}, its task-specific instantiation for vector search. By coupling geometric similarity with spectral information, the proposed method improves head--tail coherence and semantic alignment relative to purely geometric retrieval methods. It further supports adaptive search behavior through $\tau$-modulation, providing the flexibility increasingly required by modern Retrieval-Augmented Generation (RAG) pipelines. We present the complete algorithmic pipeline, establish its theoretical foundation through \texttt{epiplexity}, and evaluate the approach across benchmark and industrial settings using the open-source \texttt{arrowspace} library.
\end{abstract}


\section{Introduction}
\label{sec:introduction}
Vector retrieval systems overwhelmingly rely on cosine similarity, inner product,
or related vector-matrix product scores as their primary ranking signal
~\cite{VecDBsurvey, simhadri2024results}. Given a query and a corpus item, cosine
similarity measures their angular proximity in the ambient embedding space, but it
does not explicitly account for the statistical and topological structure of the
corpus from which those embeddings are drawn~\cite{steck2024cosine,kuffo2026semanticrecall}.
This creates a descriptive limitation: items may be geometrically close while
belonging to different regions of the feature manifold, whereas geometrically
distant items may still share structural affinity through the corpus topology.

This limitation is especially visible in the tail of ranked retrieval results,
where weakly grounded or noisy evidence can affect downstream applications (i.e Retrieval-Augmented
Generation systems)~\cite{bruckhaus2024rag,ibm2025rag}. A result set with
high average cosine similarity may still be \emph{topologically scattered}: its
items can originate from or induce poorly connected corpus communities; therefore providing the language model with a cluttered context despite strong
local geometric scores. 
Similarly, out-of-distribution (OOD) queries receive no
explicit warning signal from cosine scoring alone: a query that falls in a
spectrally irregular or high-energy region of the feature space can be ranked in
the same way as a well-grounded in-distribution query~\cite{aumuller2021role}.
We refer to this missing structural component as a \emph{semantic gap}: a
corpus-level signal that is not directly encoded by pairwise geometric similarity
but is relevant for semantic retrieval, this gap can be addressed by semantic search leveraging spectral information~\cite{moriondo2025arrowspace, moriondo2026epiplexity}.

This paper introduces \emph{SPectral INdexing} (SPIN), a
search-specific procedure that leverages the feature-space graph Laplacian
$L_F$ for retrieval and ranking that addresses the semantic gap, providing fully-featured semantic search.
\\
SPIN uses the spectral signal induced
by \emph{Graph Wiring} (GW), a
physical-network-inspired framework for topological search, to capture corpus-level structure that is not visible to geometric search.
SPIN constructs a feature-space graph (as computed by GW) to expose spectral structure from the embedding corpus. Beside geometric similarity, SPIN augments it with a discrete topological signal (spectral signal) that can be efficiently computed~\cite{shuman2013emerging,dong2016learning,sandryhaila2013discrete}.

The theoretical foundation of GW is supported through \textbf{epiplexity}, an
information-theoretic framework showing that the graph Laplacian constructed by GW
carries structural information. Following~\cite{finzi2026epiplexity}, where we use a
two-part Minimum Description Length (MDL) criterion, together with independent structural diagnostics, to prove that the constructed spectral representation encodes information beyond the original geometric scores (details in~\cite{moriondo2026epiplexity}).

The claims that we are supporting are:
\begin{itemize}
    \item To have a fully-featured semantic search it is needed to augment geometric search with spectral information, so to address the semantic gap;
    \item SPIN addresses this problem by-design providing the semantic lift needed, by leveraging the structural information made evident by epiplexity.
\end{itemize}

Our central claim is that constructing a graph Laplacian over the feature-space graph, and deriving an energy-dispersion signal from this structure,
encodes semantic correlations that are not considered by geometric search alone. Empirically, this spectral structure enables retrieval modes that preserve embedding-neighborhood affinity while producing more coherent ranked lists.\\
We evaluate GW and SPIN on retrieval and ranking tasks over two datasets: \textbf{CVE\texttrademark{} 1999--2026}, an industry-standard corpus used in our first experiment; \textbf{TREC-COVID}, a scientific retrieval benchmark with biomedical queries and human relevance judgments.

On  CVE\texttrademark{} experiment~\scref{cve_case_study}, SPIN improves ranking consistency and semantic coherence compared to cosine retrieval, producing flatter and more stable score tails with up to \(29.9\%\) lower Tail CV and \(29.8\%\) lower Tail Decay, while preserving near-perfect Tolerant Recall (\(0.993\)). SPIN also consistently improves semantic uplift across most queries (avg. \(+0.573\)).
On the human-labeled TREC-COVID~\scref{trec_covid_validation}, SPIN improves retrieval quality over cosine retrieval, achieving gains of approximately \(2.8\%\) in Relevance Recall and \(2.6\%\) in NDCG. Beyond relevance metrics, SPIN also substantially improves semantic coherence, increasing semantic uplift from \(0.00\) under cosine retrieval to \(0.334\), indicating that feature-space topology helps retrieve more semantically coherent neighborhoods beyond pure geometric similarity.
Finally, more generally, the epiplexity-based analysis~\cite{pyarrowspace2026} suggests
that GW can expose additional structural information beyond purely geometric retrieval signals, opening the way to fully-fledged semantic search.

\section{Preliminaries}
\label{sec:preliminaries}
We introduce the spectral and information-theoretic concepts underlying SPIN.

\paragraph{Graph Laplacian}
\label{spectral_graph}
Classical Spectral Graph Theory has been developed extensively to be used on the item-space \(N \times F\):  given a weighted undirected graph \(G = (V, E, W)\) with adjacency matrix \(A_{w_{i,j}} \in \mathbb{R}^{N \times N}\) and degree matrix \(D\), the \emph{combinatorial Laplacian} is \(L = D - A\), a symmetric positive
semi-definite matrix whose eigenvalues \(0 = \lambda_1 \leq \lambda_2 \leq \cdots
\leq \lambda_n\) encode the connectivity structure of the graph; the
\emph{normalized Laplacian} $\mathcal{L} = D^{-1/2} L D^{-1/2}$ scales spectral
to $[0, 2]$ for heterogeneous degree distributions.

\paragraph{Information-theoretic foundations: MDL and epiplexity}
\label{epiplexity}
According to \cite{finzi2026epiplexity}, classical Shannon entropy \(H(X) = -\sum_x p(x) \log_2 p(x)\) conflates structural regularity with irreducible randomness---adding independent noise and
discovering a new pattern both increase \(H(X)\) identically---whereas
\emph{epiplexity} resolves this by
conditioning on a computational budget \(T\). For a \(T\)-time probabilistic model
$P^* = \arg\min_{P \in \mathcal{P}_T} \bigl[|P| + \mathbb{E}[\log_2 1/P(X)]\bigr]$,
the \emph{structural information} $S_T(X)$ (model bits) captures learnable
regularity, while the \emph{random information} $H_T(X)$ (data bits) captures irreducible
per-sample uncertainty, so that the
two-part MDL code satisfies $\mathrm{MDL}_T(X) = S_T(X) + \sum_i H_T(x_i)$
and the compression test $\mathrm{MDL}_T(X) < N \cdot F \cdot b$ constitutes a
formal certificate that $P^*$ has extracted non-trivial structure
(\cite{vitanyi2000mdl, grunwald2007mdl}). Raising $T$ reveals more structure,
increasing $S_T$ and decreasing $H_T$ (see details ~\cite{moriondo2026epiplexity}). 


\section{Spectral Indexing: search via the feature-space Laplacian} \label{spin_definition}
To address the semantic gap between the embedding model's semantics and geometric search's lack of semantics handling, we need a tool to access the semantic information in the vector space.
For this end we propose SPIN, a novel application based on what is defined in \scref{spectral_graph}.

\paragraph{The \textit{feature-space} graph Laplacian} GW is a variant of graph Laplacian computation where the core variation is in the use of the feature-space~\cite{spielman2007spectral, luxburg2007tutorial}. For this end, we compute a coarser representation of the dataset \(C \times F << N \times F \) to efficiently compute the graph Laplacian, where the optimal number of clusters \(C\) is computed via the Johnson–Lindenstrauss lemma~\cite{johnson1984extensions}. 
From this \(C \times F \) space, we take the transpose \(F \times C \) and compute the graph Laplacian \(L_F \in \mathbb{R}^{F \times F} \) as in \ref{spectral_graph}.

 
\paragraph{The Rayleigh quotient as spectral smoothness}
Once \(L_F\) is available we build a dispersion network: \(L_F\) can be used to compute, for each element \(x\) of the space and any other query vector, a per-item Rayleigh quotient~\cite{rayleigh1877sound}
\[
R(x) = \frac{x^\top L_F x}{x^\top x}
\]
as normalized Dirichlet energy~\cite{evans1998partial}; physical interpretation as ``potential energy per unit mass'' on the feature manifold (\scref{appendix_rayleigh_dirichlet}). This characterizes the vector spaces in topological terms, high energy (peaks, rough) and low energy (hills, smooth).


 
\paragraph{Bounded Spectral Indexing}
The spectral distribution on all the elements in the space that is computed from the Rayleigh quotients is named \textit{taumode}. It is a sequence with each bounded element computed as
\[
\lambda^x_{\tau} = \frac{R(x)}{R(x) + \epsilon} \in [0,1)
\]
dispersion (Gini-like) term. That produces for each vector \(x\) a synthetic score that is the actual \(\lambda_\tau\) used as a positional 1D-score synthesizing the position of the vector in the space; at each query \(q\) the similarity is computed: \(sim_\lambda =  \lambda_\tau^q - \lambda_\tau^i\) .

\subsection{SPIN: lambda-aware similarity search}
\label{definition}

SPIN computes all the lambdas for the vector space to rank vectors from the most connected (low Rayleigh) to the less connected (high Rayleigh).

This algorithm as defined allows the linear combination of geometric (cosine) similarity with the spectral similarity:
\[
dist_{SPIN} = \alpha \cdot sim_{\cos}(q,i) + (1 - \alpha) \cdot sim_\lambda(q,i)
\]
SPIN is an augmentation for geometric similarity: the first term is the geometric component; the second term is the spectral component that acts as a differentiator of the geometric signal. Outliers are denoted by high spectral energy, so if the geometric distance is picking a false neighbor in a different community the spectral distance corrects the measurement. 
The \(\alpha\) in this linear combination is named the \textbf{tau} of the search in SPIN as it allows the modulation of how much spectral signal (and topological, as the graph Laplacian is a discrete topological representation) is embedded in the similarity search. Retrieval with cosine-only is \(\alpha = 1.0\). \textit{taumode} is the distribution generated by this convex map between cosine and spectral that allows to evaluate proximity in terms of raw distance but also in term of approximate topology. For example, the value of \textit{tau} that maximizes search quality for the CVE™ dataset is \(tau=0.42\), so the presence of 58\% of spectral signal provides relevant augmentation on geometric-only signal as discussed in \scref{evaluation}. 

\paragraph{Energy networks} Use of heat kernel with the Laplacian's random-walk form (\(L = I - D^{-1} A\)) on the item-space is described in \cite{chung2007heatkernelpagerank} and heat kernels have been used in graph characterization  in \cite{bai2007heatkernel}. \textit{GW's novel foundational idea is to apply an energy network interpretation to the normalized Laplacian in \textbf{feature-space to reconstruct semantics} encoded in the input vector space} (e.g. treating feature-space as the semantic web or metadata of the space).\\
A semantically typical item produces a \emph{smooth} (low-energy) signal on \(L_F\): adjacent feature nodes receive similar activation values. Conversely, an anomalous or ODD item (\emph{rough}, high-energy) (\cite{shuman2013emerging, sandryhaila2013discrete})---information that geometric search cannot access (more in \ref{appendix_rayleigh_dirichlet}). Computing the difference \(sim_\lambda(q,i)\) rewards items that are not merely close to the query \(q\) in angle but that also occupy the same energy region of the feature manifold in which \(i\) resides. Cosine acts as high-locality signal while \(sim_\lambda(q,i)\) acts as a corpus-structure signal. These principles allow all the capabilities described in this paper that are embedded in SPIN by-design.

\section{Experimental evaluation}
\label{evaluation}
\subsection{Setup} 
The experiments have been run on an Apple Silicon M4 laptop with 24Gb of RAM. The embedding model has been fined-tuned locally as well.
\subsection{Datasets}
\label{datasets}
We evaluate SPIN on two retrieval benchmarks. The first is the CVE™ vulnerability corpus~\cite{mitre2026cve}, a cybersecurity retrieval workload comprising \(347{,}113\) vulnerability reports. The evaluation uses \(50\) LLM-generated vulnerability queries designed to emulate realistic semantic search scenarios in cybersecurity analysis. The second benchmark is TREC-COVID~\cite{voorhees2021treccovid}, a biomedical retrieval dataset derived from the CORD-19 scientific literature collection, containing approximately \(171{,}000\) scientific articles and \(50\) expert-designed search topics with graded relevance judgments. TREC-COVID provides expert-annotated graded relevance labels, enabling relevance-aware evaluation using metrics such as NDCG.

\paragraph{Implementation}
We provide an Open Source implementation for GW and SPIN in an unified library \texttt{arrowspace} in both Rust and Python~\cite{pyarrowspace2026}.
\texttt{arrowspace} exposes several hyperparameters controlling the construction of the feature-space graph: \(\varepsilon\), the neighborhood radius threshold; \(k\), the maximum neighborhood size; \(p\), the exponent used in adjacency-weight computation; and \(\sigma\), the bandwidth parameter of the Gaussian kernel defining adjacency weights.

\paragraph{Retrieval evaluation}
\label{retrieval_evaluation}
Classical retrieval metrics remain largely insensitive to the topological structure of the embedding manifold~\cite{steck2024cosine,kuffo2026semanticrecall}. Since retrieval failures in domain-specific corpora are closely linked to context degradation and hallucination~\cite{magesh2025hallucination}, we complement standard effectiveness metrics with graph-signal-aware evaluation. Specifically, we report: (i) \textbf{Spearman's $\rho$} and \textbf{Kendall's $\tau$} to measure the divergence between cosine and $\lambda_\tau$-aware rankings; (ii) \textbf{Relevance Recall@$k$} and \textbf{NDCG@$k$}, to verify that SPIN improves the fraction of retrieved documents judged relevant by the TREC-COVID qrels while preserving graded relevance quality. (iii) \textbf{head--tail metrics} (\textit{T/H Ratio}, \textit{TailCV}, and \textit{TailDecay}) to characterize structural coherence and score stability in the lower-ranked results where cosine retrieval degrades most severely; and (iv) \textit{Traditional Recall}, \textit{Tolerant Recall}, and \textit{Semantic Recall}~\cite{kuffo2026semanticrecall}, together with \textit{Semantic Uplift}, to evaluate semantic robustness beyond exact geometric matching. Detailed definitions of all evaluation metrics are provided in Appendix~\ref{retrieval_metrics_appendix}.

\paragraph{Procedure}
Documents and queries are encoded using a domain-adapted Sentence-BERT embedding model~\cite{reimers-2019-sentence-bert}, with a lightweight fine-tuning step performed on a small subset of the target corpus (\(3\text{--}5\%\) of the data). To improve graph formation in the feature-space, vectors are uniformly scaled by a factor of \(1.12\), preserving geometric relations while mitigating normalization effects. We compare three retrieval configurations: cosine retrieval with \(\tau = 1.0\), hybrid retrieval with \(\tau = 0.72\), and \texttt{taumode} retrieval with \(\tau = 0.42\). The parameter \(\tau\) controls the contribution of the spectral signal in SPIN, where \(\tau=1.0\) reduces to pure cosine retrieval and lower values progressively increase the influence of spectral information. Graph-construction and indexing hyperparameters are tuned using Pareto-optimal trade-offs between retrieval quality and structural coherence. All experimental scripts and parameters are released for reproducibility~\cite{pyarrowspace2026}.

\subsection{CVE\texttrademark{} experiment}
\label{cve_case_study}
We evaluate SPIN on CVE dataset across three spectral modes: cosine (\textit{tau} = 1.0), hybrid (\textit{tau} = 0.72) and taumode (\textit{tau} = 0.42). Cosine produces standard geometric search results under cosine metric similarity, while as we increase \textit{tau}, the search produces flatter, more uniform and slower-decaying score tails while preserving high Tolerant Recall.






\paragraph{Tail-shape metrics}
Table~\ref{tab:tail} summarizes the \(K_h=3\) tail metrics. taumode has the highest T/H Ratio, lowest Tail CV, and lowest Tail Decay. This is evidence of more balanced tail, where the semantic affinity of top-ranks show smaller difference with the low-ranks compared to cosine.

\begin{table}[h]
\centering
\caption{Mean tail-shape metrics at \(K_h=3\).}
\label{tab:tail}
\begin{tabular}{lrrrrr}
\toprule
Method & Head mean & Tail mean & T/H Ratio & Tail CV & Tail Decay \\
\midrule
Cosine \((\tau=1.0)\) & 0.843 & 0.831 & 0.98600 & 0.00396 & 0.000512 \\
hybrid \((\tau=0.72)\) & 0.875 & 0.863 & 0.98639 & 0.00382 & 0.000509 \\
taumode \((\tau=0.42)\) & \textbf{0.924} & \textbf{0.915} & \textbf{0.99089} & \textbf{0.00254} & \textbf{0.000359} \\
\bottomrule
\end{tabular}
\end{table}

\begin{table}[h]
\centering
\caption{Per-query winners over 50 queries}
\begin{tabular}{lrrr}
\toprule
Metric & Cosine wins & hybrid wins & taumode wins \\
\midrule
T/H Ratio & 5 & 2 & \textbf{43} \\
Tail CV & 6 & 2 & \textbf{42} \\
Tail Decay & 8 & 2 & \textbf{40} \\
\bottomrule
\end{tabular}
\end{table}

\begin{table}[h]
\centering
\caption{Recall metrics.}
\label{tab:recall}
\begin{tabular}{lrrr}
\toprule
Recall variant & Cosine & hybrid & taumode \\
\midrule
Traditional Recall@25 & 1.000 & \(0.518\pm0.392\) & \(0.420\pm0.419\) \\
Semantic Recall@25 & 1.000 & \(0.554\pm0.417\) & \(0.423\pm0.431\) \\
Tolerant Recall@25 & 1.000 & \(0.954\pm0.184\) & \(\mathbf{0.993\pm0.050}\) \\
\bottomrule
\end{tabular}
\end{table}

\begin{figure}[!htp]
  \centering
    \includegraphics[width=0.8\linewidth]{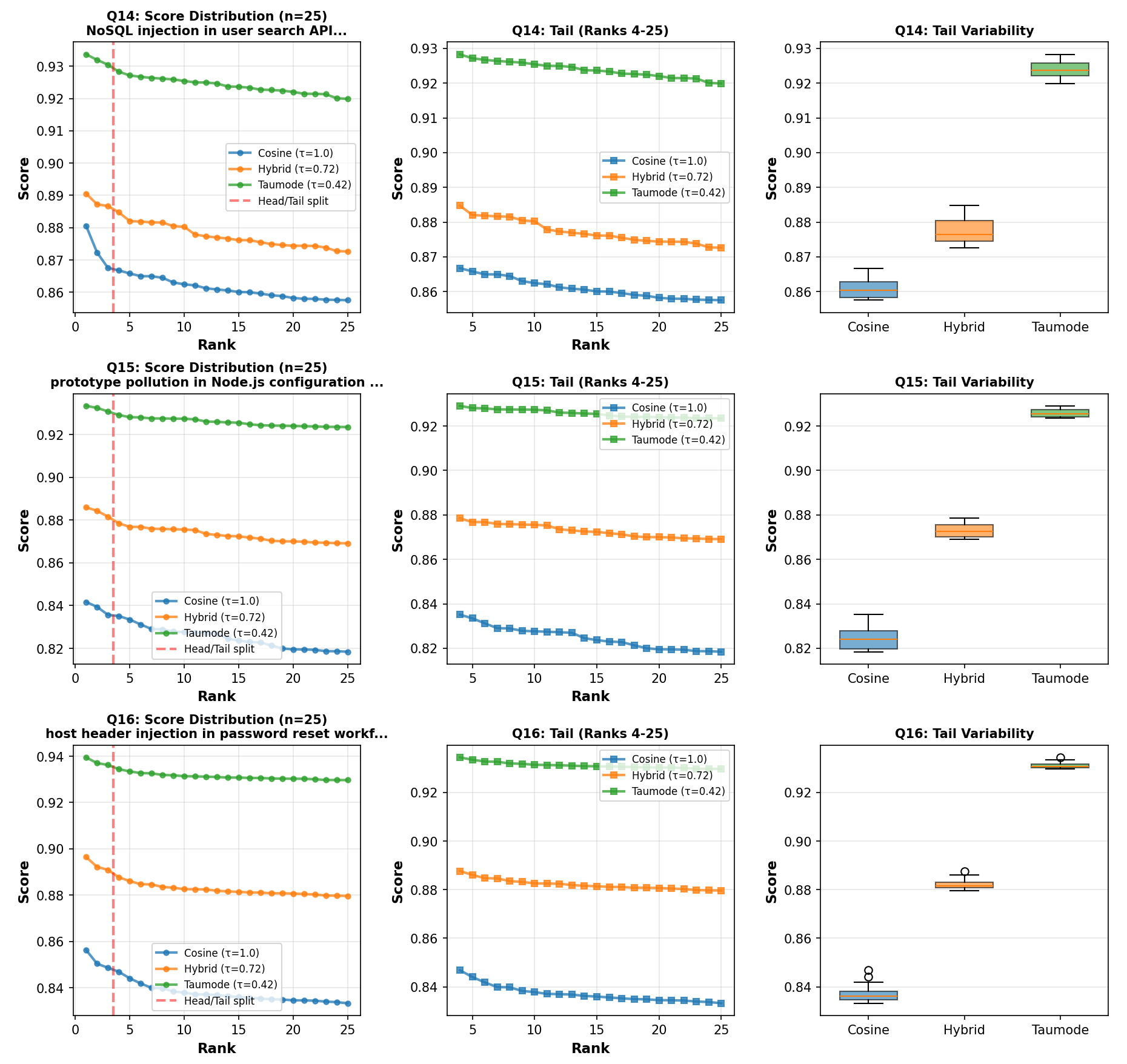}
    \caption{Tail quality for 4 queries for all tested modes.}
    \label{fig:exp_1_tail_quality}
\end{figure}

\paragraph{Head--\(K_h\) sweep.}
Head--\(K_h\) is tested in a comparative sweep across \(K_h \in \{3, 5, 10\}\),
where the head--tail boundary is progressively widened.
The results confirm that the taumode advantage is stable across all tested head lengths (additional results in table \ref{tab:head_k})
The tail-to-head ratio and tail coefficient of variation both improve marginally for all three methods as \(K_h\) increases as visible in fig. \ref{fig:head_k_sweep}.
taumode also consistently exhibits a lower tail decay rate and a tighter
tail score distribution (lower \(\mathrm{CV}\)).
Overall, the sweep demonstrates that the observed ranking improvement is a robust property of the \textit{tau} reranking  mechanism and not an artifact of any particular choice of head--tail boundary. Fig. \ref{fig:exp_1_tail_quality} visualizes the reduced delta from the top-ranking results and the low-ranking results by highlighting a less steep slope in the scores curve between head and tail (left and middle columns). On the right column we can see that the scores' tail variability is reduced in the case of hybrid and taumode; this implies that taumode returns better scores in all rankings but also more balanced scores among tail's elements.

\paragraph{Recall and ranking agreement}

Table~\ref{tab:recall} summarizes recall variants. Cosine is takes as a reference but with the limitation described in \ref{circular_cosine}, therefore trivially scores 1.0 in Traditional Recall; Tolerant Recall is the more informative non-circular measure for taumode as it shows that even embedding spectral information keeps the metric still very close to the reference geometric score.

\begin{figure*}[!htp]
    \centering

    \begin{subfigure}[t]{0.32\linewidth}
        \centering
        \includegraphics[width=\linewidth]{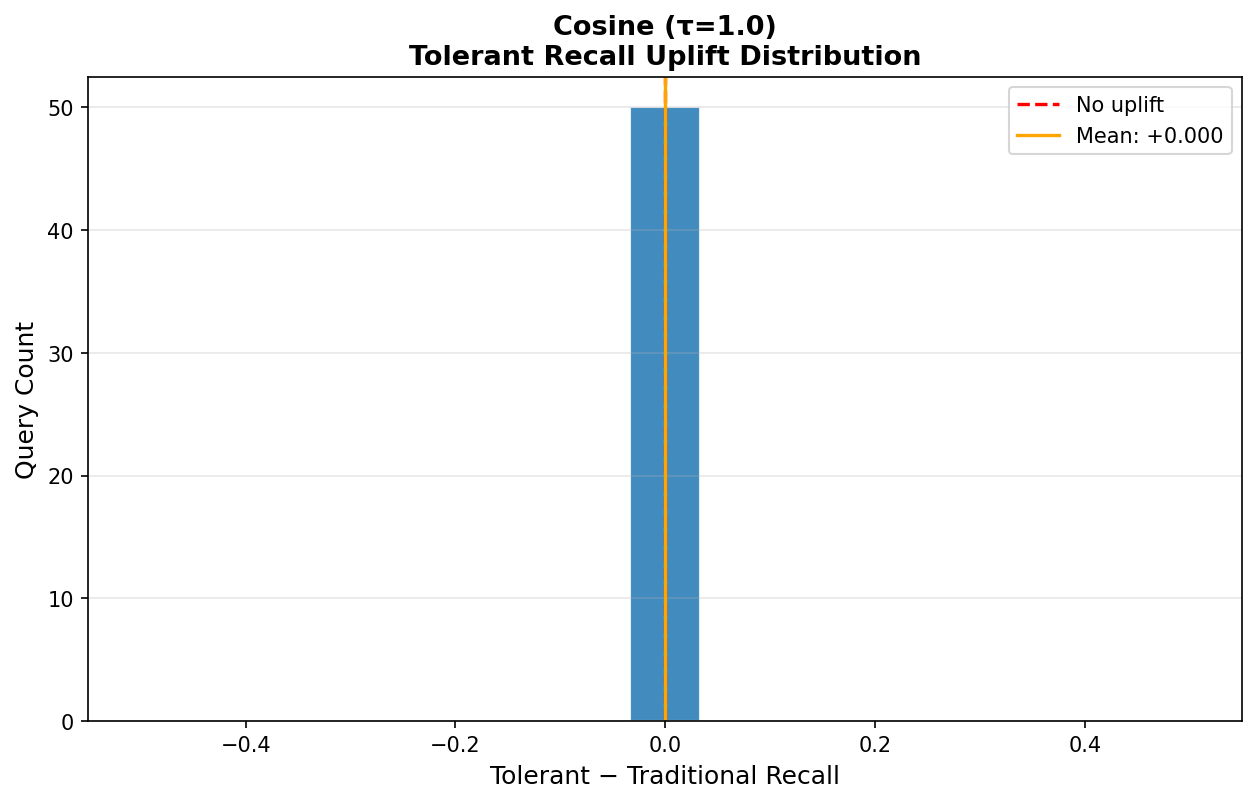}
        \caption{Cosine retrieval}
        \label{fig:cve:semantic_uplift_cosine}
    \end{subfigure}
    \hfill
    \begin{subfigure}[t]{0.32\linewidth}
        \centering
        \includegraphics[width=\linewidth]{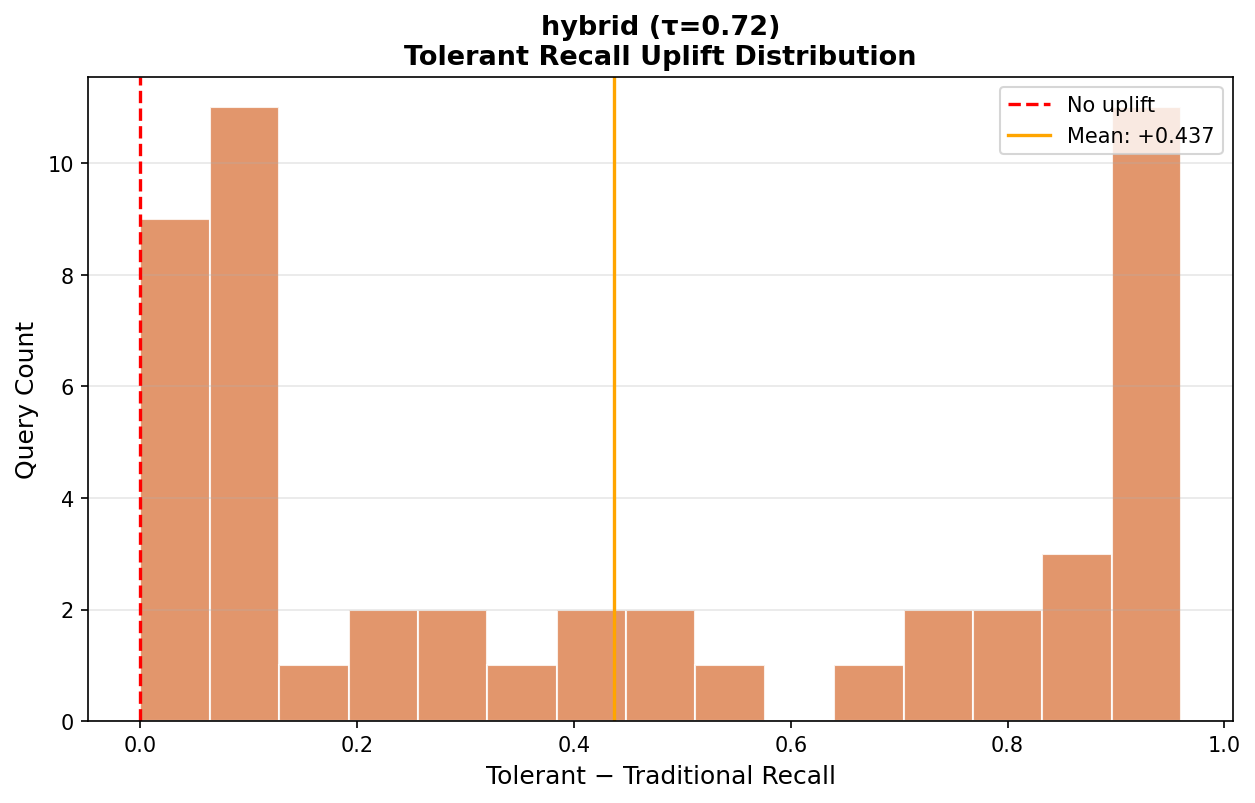}
        \caption{Hybrid retrieval}
        \label{fig:cve:semantic_uplift_hybrid}
    \end{subfigure}
    \hfill
    \begin{subfigure}[t]{0.32\linewidth}
        \centering
        \includegraphics[width=\linewidth]{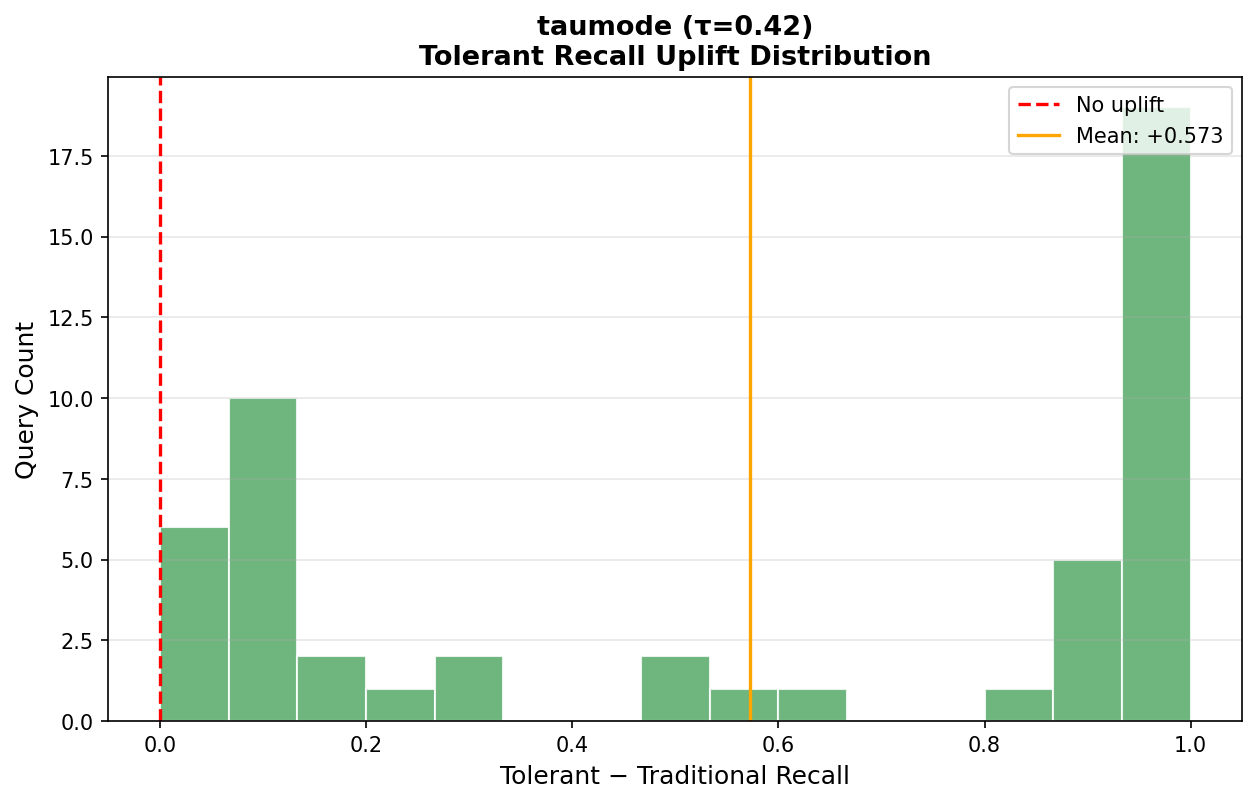}
        \caption{Taumode retrieval}
        \label{fig:cve:semantic_uplift_cve}
    \end{subfigure}

    \caption{\textit{CVE\texttrademark{} dataset: Semantic Uplift}. Cosine (left) provides no uplift, taumode (right) provides relevant uplift for most of queries}
    \label{fig:cve:semantic_uplift}
\end{figure*}

\paragraph{Semantic uplift.} \label{semantic_uplift}
Semantic uplift is defined as the per-query difference
\(R_{\mathrm{uplift}} = R_{\mathrm{tol}} - R_{\mathrm{trad}}\).
The data shows that taumode exhibits notable positive uplift compared to cosine on several queries (fig. \ref{fig:cve:semantic_uplift_cve}).
The Tolerant Recall ($R_{\mathrm{tol}}$) is specifically designed to soften this reference dependency: across all negative-uplift queries, taumode achieves $R_{\mathrm{tol}} = 1.00$, confirming that no diverging document is irretrievable within a modest score relaxation.
A qualitative case-by-case inspection of these queries is available in Appendix~D \ref{appendix_semantic_qual} for the
reader's judgment.
For a more formal assessment against an independently labeled dataset,
the second experiment is presented in Section \scref{trec_covid_validation} that confirms this interpretations on the negative uplift queries. \par

\subsection{TREC-COVID experiment}
\label{trec_covid_validation}
We evaluate SPIN on the TREC-COVID benchmark~\cite{voorhees2021treccovid}, a labeled document-queries COVID-related dataset containing expert-designed queries, and human-annotated relevance judgments. Unlike standard ANN benchmarks that only provide metric nearest-neighbor ground truth, TREC-COVID enables the evaluation of semantic relevance using human labels.
The experiments use the official TREC-COVID queries and labels to assess whether SPIN retrieves documents considered relevant by human annotators. We report NDCG and Relevance Recall on K=10, providing a more realistic assessment for downstream retrieval applications such as RAG and scientific search.

\begin{figure}[!htp]
    \centering

    \begin{subfigure}[b]{0.48\linewidth}
        \centering
        \includegraphics[width=\linewidth]{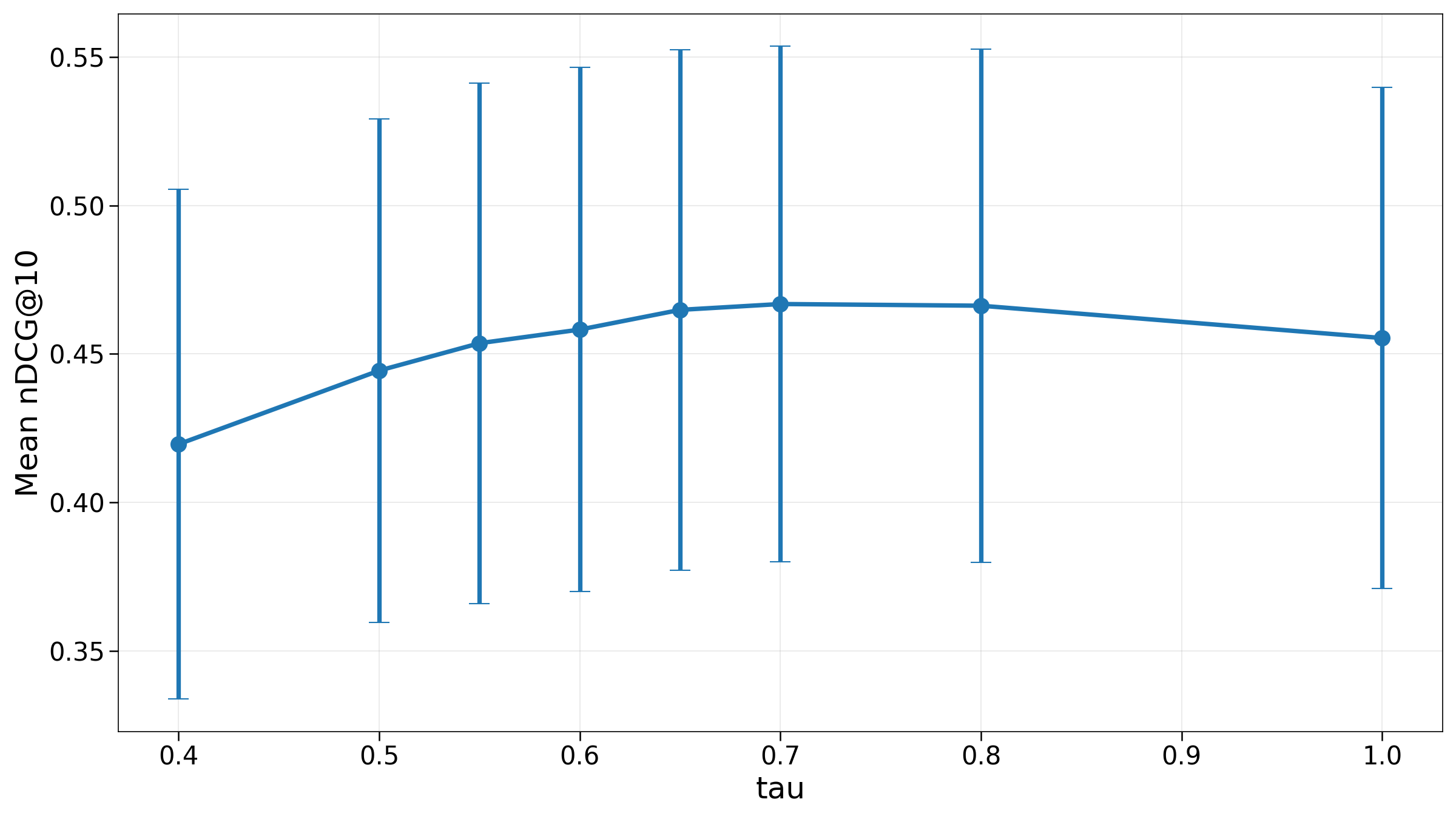}
        \caption{NDCG across different \textit{tau} values.}
        \label{fig:trec:ndcg_tau}
    \end{subfigure}
    \hfill
    \begin{subfigure}[b]{0.48\linewidth}
        \centering
        \includegraphics[width=\linewidth]{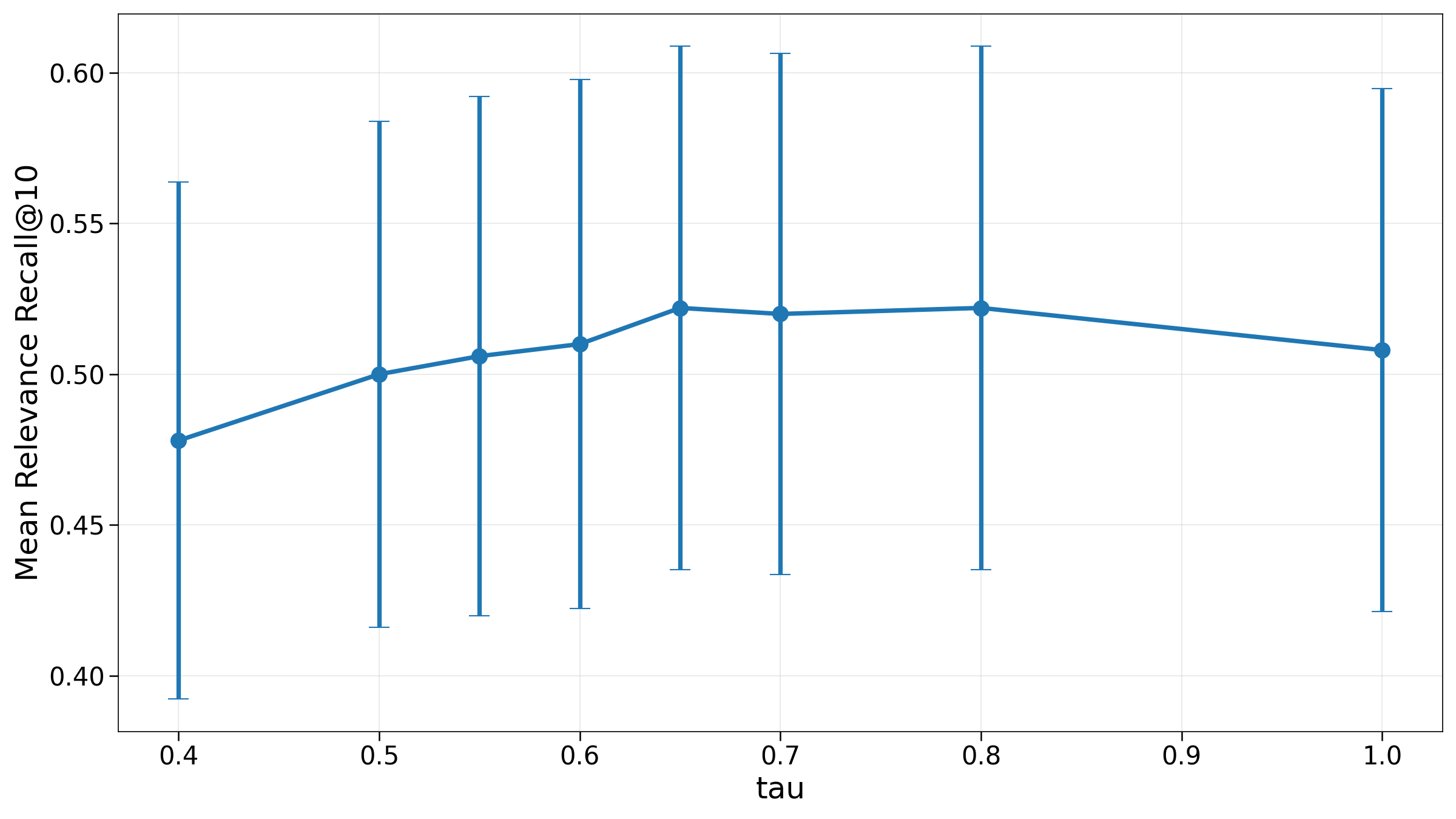}
        \caption{Relevance recall across different \textit{tau} values.}
        \label{fig:trec:relevance_recall_tau}
    \end{subfigure}

    \caption{Impact of the spectral mixing parameter \textit{tau}  on retrieval quality. Lower \textit{tau} values incorporate stronger spectral information, while \textit{tau}=1 corresponds to pure cosine similarity.}
    \label{fig:trec:tau_sensitivity}
\end{figure}

\paragraph{Retrieval Quality} Figure~\ref{fig:trec:relevance_recall_tau} shows that incorporating spectral information improves Relevance Recall@10 compared to pure cosine similarity (tau=1.0). Cosine reaches a mean recall of $0.508$, while intermediate $\tau$ values achieve the best performance around $tau \in [0.65,0.80]$, peaking near $0.522$. This indicates that combining geometric similarity with feature-space topology helps retrieve more relevant documents. Very low tau values ($0.40$) slightly degrade performance, suggesting that excessive spectral influence weakens semantic alignment. Despite noticeable query-level variability, the gains remain stable across neighboring $tau$ values.

Figure~\ref{fig:trec:ndcg_tau} reports the corresponding NDCG@10 results, where the impact on ranking quality becomes more pronounced. Pure cosine similarity achieves $0.455$, while the best hybrid settings ($tau=0.70$ and $0.80$) reach approximately $0.467$. Since NDCG accounts for both ranking order and graded relevance, the improvement suggests that spectral information helps rank highly relevant documents earlier. Performance increases progressively from $ttau=0.40$ to $tau\approx0.70$, then saturates, indicating that moderate spectral integration provides the most effective balance between semantic geometry and corpus topology.

\paragraph{Tail-shape metrics} Table~\ref{tab:trec:tail_metrics} reports the tail-shape statistics of the ranked lists on TREC-COVID at $k=10$. Compared to cosine retrieval, SPIN progressively improves both head and tail relevance as spectral information is incorporated. In particular, \textit{taumode} ($tau=0.4$) achieves the highest Tail/Head ratio ($0.98$), indicating that relevance remains nearly as strong at the tail of the ranking as at the top positions. At the same time, the lower Tail CV and Tail Decay values suggest a more stable and less rapidly degrading ranking distribution. These results indicate that spectral retrieval not only improves top-ranked relevance but also preserves semantic consistency deeper in rankings.
\begin{table}[h]
\centering
\caption{Mean tail-shape metrics at $k=10$ from TREC logs.}
\label{tab:trec:tail_metrics}
\begin{tabular}{lrrrrr}
\toprule
Method & Head mean & Tail mean & T/H Ratio & Tail CV & Tail Decay \\
\midrule
Cosine (tau=1.0) & 0.802 & 0.761 & 0.94862 & 0.01362 & 0.004203 \\
hybrid (tau=0.7) & 0.849 & 0.820 & 0.96616 & 0.00886 & 0.002964 \\
taumode (tau=0.4) & 0.908 & 0.890 & \textbf{0.98009} & \textbf{0.00459} & \textbf{0.001647} \\
\bottomrule
\end{tabular}
\end{table}

\begin{figure*}[!htp]
    \centering

    \begin{subfigure}[t]{0.32\linewidth}
        \centering
        \includegraphics[width=\linewidth]{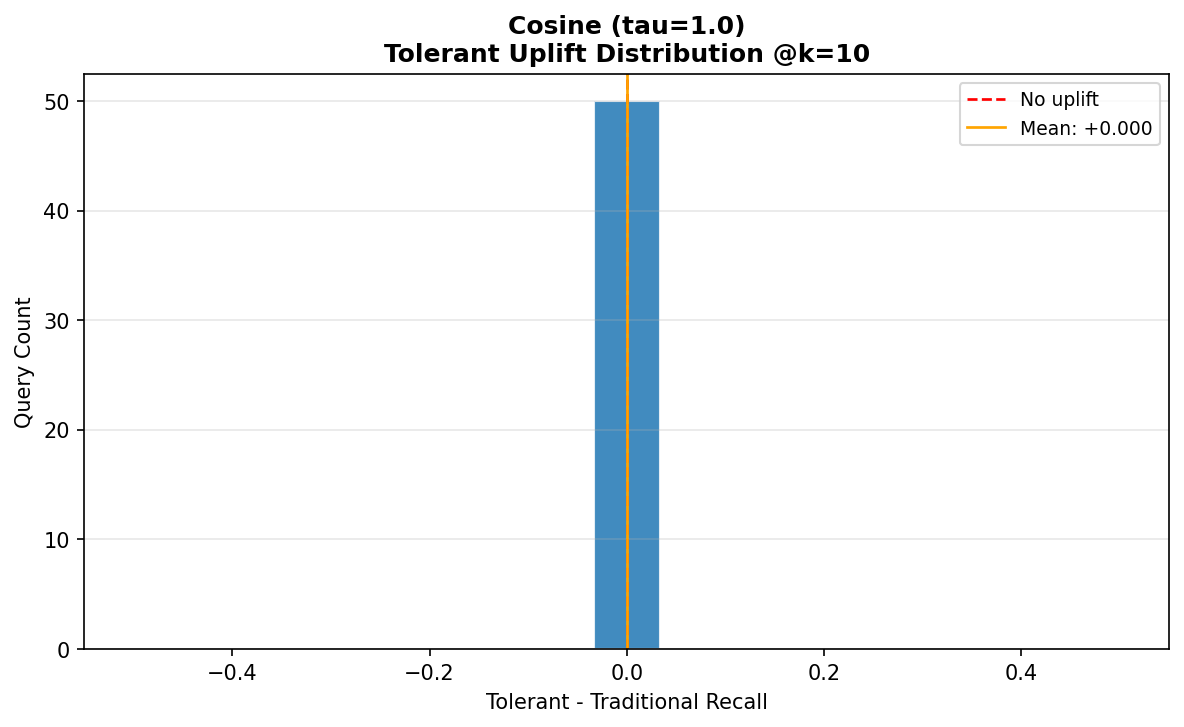}
        \caption{Cosine retrieval}
        \label{fig:trec:semantic_uplift_a}
    \end{subfigure}
    \hfill
    \begin{subfigure}[t]{0.32\linewidth}
        \centering
        \includegraphics[width=\linewidth]{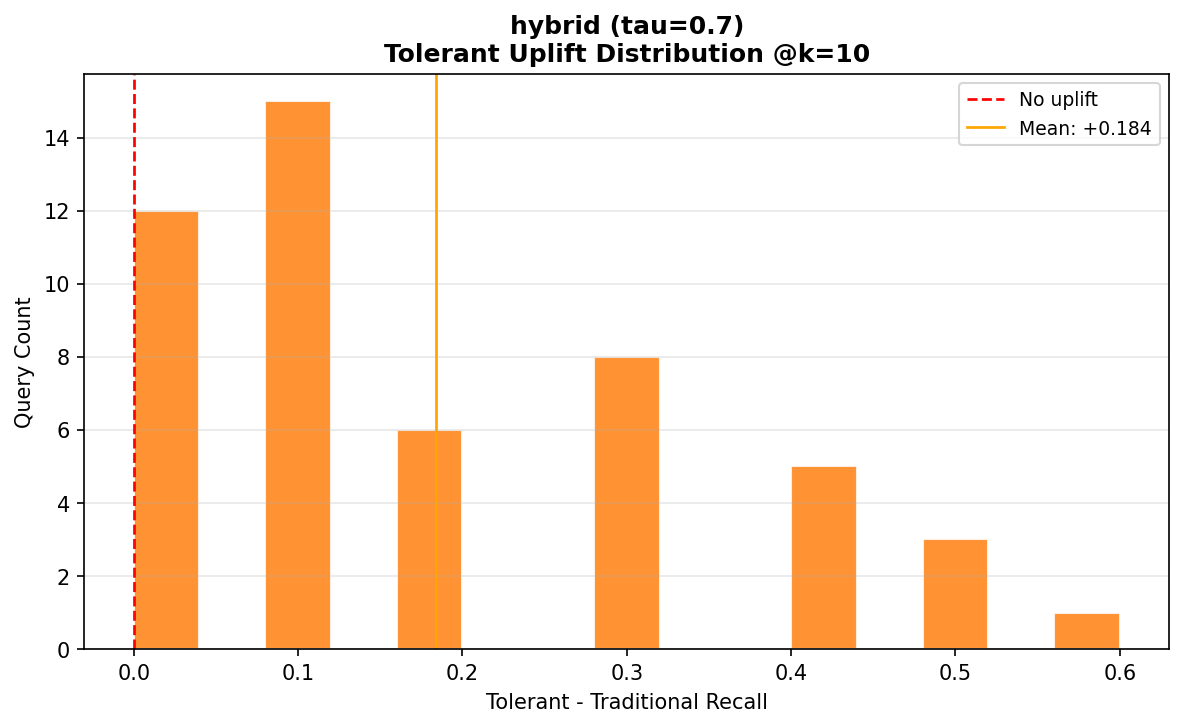}
        \caption{Hybrid retrieval}
        \label{fig:trec:semantic_uplift_b}
    \end{subfigure}
    \hfill
    \begin{subfigure}[t]{0.32\linewidth}
        \centering
        \includegraphics[width=\linewidth]{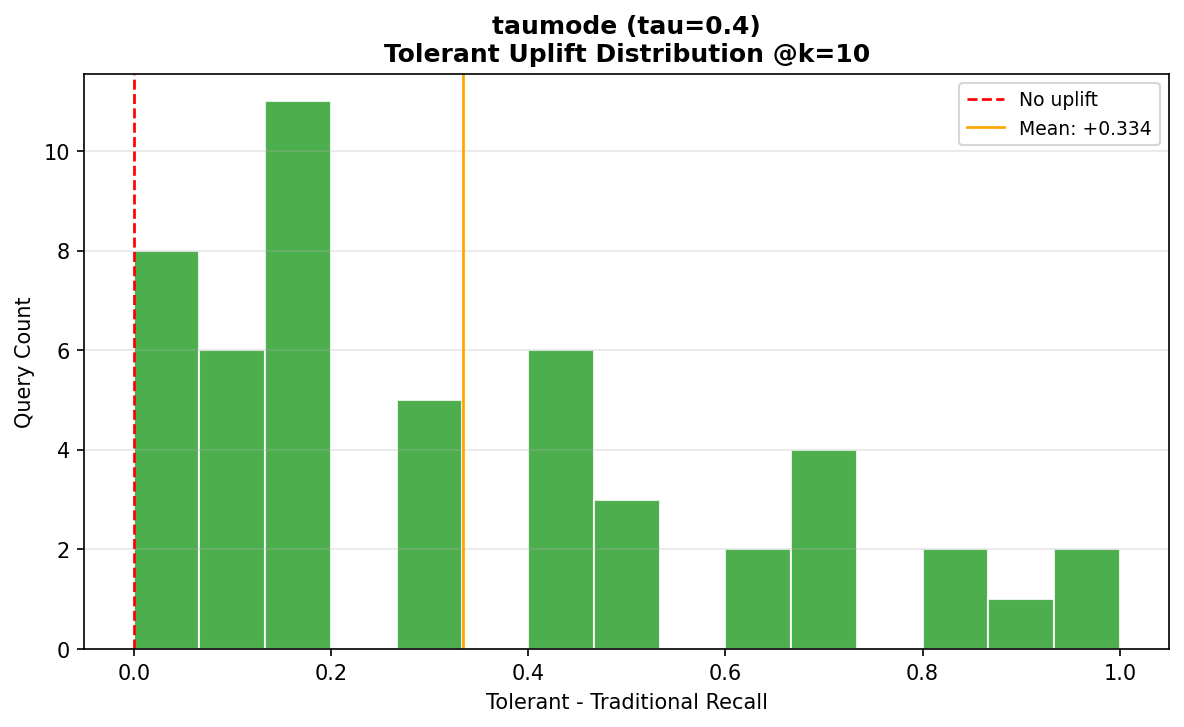}
        \caption{Taumode retrieval}
        \label{fig:trec:semantic_uplift_c}
    \end{subfigure}

    \caption{\textit{Semantic Uplift} on the Trec-Covid dataset. Cosine retrieval provides limited uplift, while the spectral retrieval modes improve semantic recall across most queries.}
    \label{fig:trec:semantic_uplift}
\end{figure*}

\paragraph{Semantic Uplift} Figure~\ref{fig:trec:semantic_uplift} reports the distribution of semantic uplift scores across the 50 TREC-COVID queries. As expected, cosine retrieval (\(\tau=1.0\), Figure~\ref{fig:trec:semantic_uplift_a}) produces no semantic uplift, with all queries concentrated at \(0.0\) and a mean uplift of \(0.00\). Introducing spectral information in the hybrid configuration (\(tau=0.7\), Figure~\ref{fig:trec:semantic_uplift_b}) shifts the distribution toward positive uplift values, yielding a mean uplift of \(0.184\). Most queries receive moderate gains between \(0.1\) and \(0.4\), while only \(12/50\) queries remain at zero uplift. The strongest effect is observed with taumode retrieval (\(\tau=0.4\), Figure~\ref{fig:trec:semantic_uplift_c}), where the distribution becomes substantially broader and shifts toward higher uplift values, reaching a mean uplift of \(0.334\). In this setting, only \(8/50\) queries exhibit no uplift, while several queries achieve gains above \(0.6\), and a small subset reaches high uplift values close to \(1.0\). 

\section{Discussion}
The core idea of GW is that an item vector \(x_i \in \mathbb{R}^{N \times F}\), when blended with the graph Laplacian in the feature-space \(L_F \in \mathbb{R}^{F \times F}\), carries measurably more discriminative information than its geometric measurement in the item-space alone, as demonstrated in the terms of epiplexity framework and structural information in ~\cite{tuned_epiplexity,moriondo2026epiplexity}.

\label{exp_interpretation}
\paragraph{Experiments interpretation.} Taken together, the evidence supports the following reading: \emph{cosine similarity is a strong local baseline but a limited global semantic index}. While cosine effectively captures geometric proximity in the embedding space, it does not explicitly model the broader topological organization of the feature manifold. SPIN instead reorders the \emph{structure} of the ranked list according to the graph's spectral-topological layout, integrating both local geometry and corpus-level connectivity information.

\begin{figure}[htp]
  \centering
  \includegraphics[width=0.8\linewidth]{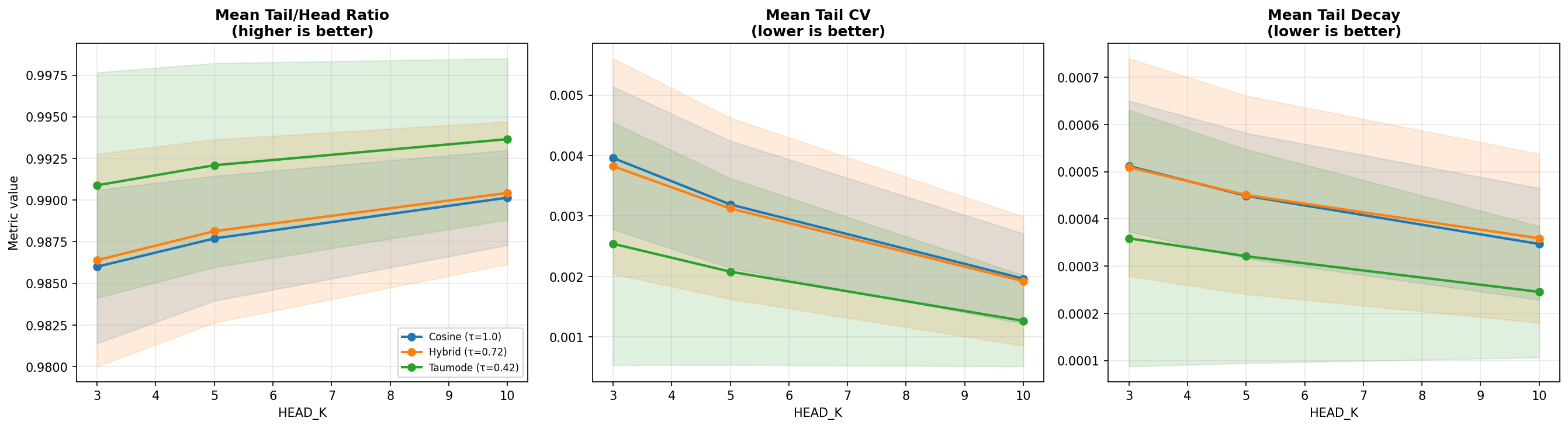}
  \caption{\textit{Head\_K} sweep: quality of taumode results improves according to tail metrics.}
  \label{fig:head_k_sweep}
\end{figure}

Across both CVE and TREC-COVID, the empirical effects are consistent. SPIN achieves near-perfect Tolerant Recall when compared to cosine, produces positive Semantic Uplift on difficult queries (see Appendix~\ref{appendix_cve_qualitative}), generates flatter and more coherent ranking tails (see Fig.~\ref{fig:exp_1_tail_quality}), and maintains stronger semantic affinity deeper into the ranking (see Fig.~\ref{fig:head_k_sweep}). On TREC-COVID, intermediate \textit{tau} values also improve both Relevance Recall and NDCG, indicating that spectral information helps recover and prioritize documents judged relevant by human assessors.

Importantly, the gains are strongest in hybrid settings rather than in purely spectral regimes, suggesting that geometric similarity and spectral topology provide complementary retrieval signals. The tail-shape analysis further indicates that SPIN preserves semantic consistency beyond the top-ranked results, reducing the rapid quality decay commonly observed with cosine-only retrieval.

Overall, these results suggest that spectral-topological retrieval improves not only local ranking quality but also the global semantic organization of the returned neighborhoods. \textit{These properties are particularly important for knowledge-intensive retrieval systems} where the objective is not only to recover exact known variants of a document, but also the broader semantic class and contextual evidence surrounding the query.

\paragraph{Computational complexity} \label{computational_costs} \texttt{ArrowSpace} algorithm in the \texttt{arrowspace} library is the current reference implementation for GW. Index construction complexity costs \(O(F \cdot \mathrm{nnz}(L_F))\), following an \(O(N \cdot F)\) incremental centroid clustering step that reduces the data matrix from \(N \times F\) to \(C \times F\). \textit{Memory/storage} requirements are \(O(F \cdot N)\) for dense features, \(O(F \cdot \mathrm{nnz}(L_F))\) for the Laplacian and \(O(N)\) for \(\lambda_\tau\) scalar index.
\texttt{ArrowSpace} is designed to keep computational cost concentrated in scalable stages: the embedding matrix is first reduced through incremental centroid clustering with \(O(N \cdot F)\) cost, after which the main spectral operations are performed on sparse feature-space. The implementation benefits from highly parallel execution on CPUs enabled by the Rayon library \cite{rayon_rust}.

\paragraph{Open problems}
\label{limitations} At current state additional attention has to be put in how the starting embeddings are generated; some precautions have to be taken when generating the vector space in terms of preserving the spectral signal and the semantics of the original context, also at graph formation time. GW and SPIN seem to work better with fine-tuned embeddings models in the case of structured documents (i.e. JSON format) but fine-tuning does not improve results for plain documents.

\subsection{Limitations}

The query set has 50 examples and shows substantial per-query variance. The score scales differ across \textit{tau}, so absolute score means should be interpreted carefully; ratio and shape metrics are more robust. The analysis uses one embedding model and one graph hyperparameter configuration, so transfer across encoders and corpora remains to be tested despite being confirmed in these two experiments.

Experiments have been run on dense vector spaces where semantical connections are well-established (text embeddings); experiments should also be run on raw sparse and numerical data to see how GW and SPIN behave. Current text embedding pipelines and vector databases are optimized for geometric search and have never considered topological/spectral signal as relevant, so no tools have been developed to leverage it. The technology has been developed in two decades; SPIN should go through the same optimization process in the next years.

\textbf{Circular cosine baseline} \label{circular_cosine} The Traditional Recall metrics use a ground truth constructed from cosine-ranked neighbors; cosine therefore attains perfect recall by construction.  This renders Traditional Recall an unreliable proxy for semantic quality and motivates the use of Tolerant Recall and Semantic Uplift as primary evaluation criteria, augmented by a head--tail analysis and labeled data-points.

\textbf{Tolerant-recall corridor width}  Tolerant Recall is computed within a fixed score tolerance band \(\epsilon\).  If \(\epsilon\) is too wide, the metric rewards any dense retrieval system; if too narrow, it penalizes semantically correct but score-distant results, that is exactly what spectral signal mitigates.
The tolerance parameter used here (\(\approx 44\%\) of each query's score range) is broad enough to capture the spectral up-scaling effect of taumode and hybrid, it has been also confirmed by the use of human annotated relevance judgments in experiment 2 (\ref{trec_covid_validation}).

\textbf{Frozen encoder} All experiments use a fixed pre-trained \texttt{SentenceTransformer} encoder~\cite{reimers2019sbert}. Since the similarity graph and spectral Laplacian are built directly from these embeddings, the spectral layer can only reorganize existing structure and cannot correct semantic biases introduced by the encoder itself. A natural extension is to jointly fine-tune the encoder on retrieval pairs (e.g., using \texttt{MultipleNegativesRankingLoss}) and rebuild the graph in the updated embedding space, allowing the representation and spectral structure to co-adapt.

\section{Conclusion and future works}
\label{conclusion}

The experiments results provide evidence that spectral indexing improves retrieval coherence. SPIN (augmenting geometric distance with spectral information) produces more accurate results in the terms explained in \ref{exp_interpretation}.
Query-by-query qualitative analysis is available in Appendix \ref{appendix_cve_qualitative} and Appendix \ref{appendix_semantic_qual}. 
Strict semantic-affinity claims are based on Semantic Recall measurements. The most defensible conclusion is, if Semantic Uplift is considered reliable, that SPIN is highly effective at producing coherent, manifold-smooth retrieval neighborhoods; this observation is further reinforced by the external relevance evaluations on TREC-COVID where improvements in both Relevance Recall and NDCG demonstrate that the spectral topological signal also translates into higher human-judged semantic relevance.

In this paper we formally defined and evaluated:
\begin{itemize}
\item \textbf{Graph Wiring(GW)}: the base algorithm, graph Laplacian in feature-space,
     which effectiveness has been defined in terms of epiplexity (structural information).
\item \textbf{SPectral INdexing (SPIN)}: the search-specific variant of GW that implements
     energy dispersion networks to achieve SOTA performance in high-dimensional semantic-heavy spaces for searching and ranking operations. 
\item \textbf{arrowspace~\cite{pyarrowspace2026}}: the implementation that has been used in the experiments. 
\end{itemize}

On the more speculative side, GW as an energy networks graph formation algorithm is a candidate as an approximate solution for the problem of surface minimization in physical networks, via a still hypothetical equivalence denoted in \cite{moriondo2026surface} (page 5). This may open research possibilities in the next years in the fields of computing physical surfaces; the same properties may hold for modeling meta-materials.
\paragraph{Future works} An hypothesis about usage of GW is that it can be used as a kind of long-term memory because the graph Laplacian is where all the invariants of the context are best represented; research about how these invariants can receive updates by new clusters in the graph is ongoing. For now, GW has relevant potential as a multi-purpose tool for datasets curation, model evaluation, models behavior predictions for data drifting (\cite{moriondo2026surface}). The implementations introduced here are examples for improving search in scenarios for which structural information can make the difference for downstream engineering applications ("spectral intelligence for data operations") like vector indexing, RAG or future classes of systems involving semantics-heavy models at scale.
\paragraph{Applications} A mix of spectral information and token-generation can provide Spectral-aware Unique Identifiers (SUIDs), an indexing tool that can drive generative retrieval \cite{moriondo2026spectraltokens}.
SPIN can be used as a new semantic skill which we call \emph{adaptive search}  for RAG and Agentic workflows where structurally inconsistent or noisy retrieved context can negatively affect downstream reasoning and generation quality. 
In this setting, the system adapts the parameter \(tau\) to the current phase of context formation. 
When the context is still being formed, the system needs more generic content to establish a broad frame, so it can rely more on geometric signal (e.g. \(tau \approx 1.0\)). 
Later, once a general frame is in place, the system often needs more specific retrieval and a finer sieve over the corpus, and \(tau\) can be reduced to embed more semantic signal (e.g., \(tau \approx 0.4\)). 
The system can also adjust \(tau\) automatically based on the state of the conversation.  We refer to this process as \emph{\(tau\)-modulation}. We believe this mechanism can mitigate problems in context formation that limit long-context applications, especially deadlocks caused by poor retrieval. 
A simple feedback loop on \(tau\) may avoid such failures; future work will explore this experimentally. 
As a vector similarity tool, SPIN makes distances between vectors semantic-aware; this improvement is measurable and reduces the gap (“semantic sink”) between the semantics captured by text embedding models and purely geometric search. 
In turn, this yields a fully-fledged semantic search suitable for knowledge-intensive retrieval. 

\section*{Acknowledgments}
The authors acknowledge the use of AI-assisted tools during the preparation of this paper, including Perplexity for literature discovery, OpenAI GPT models for drafting and result analysis, NVIDIA Nemotron for exploratory formal verification analysis, and Claude Sonnet 4.6 for coding support and implementation checks~\cite{perplexity,openai_gpt5,nvidia_nemotron,anthropic_claude46}. All technical claims, interpretations, experiments, and final editorial decisions remain the sole responsibility of the authors.

\bibliographystyle{unsrt}  
\bibliography{refs}

@article{bruckhaus2024rag,
  author = {Bruckhaus, T.},
  title = {RAG does not work for enterprises},
  journal = {arXiv preprint arXiv:2406.04369},
  year = {2024},
  url = {https://arxiv.org/abs/2406.04369}
}

@article{finzi2026epiplexity,
  author = {Finzi, M. and Potapczynski, A. and Goldblum, M. and Wilson, A. G.},
  title = {From entropy to epiplexity: Rethinking information for computationally bounded intelligence},
  journal = {arXiv preprint arXiv:2601.03220},
  year = {2026},
  url = {https://arxiv.org/abs/2601.03220}
}

@misc{ibm2025rag,
  author = {{IBM}},
  title = {RAG problems persist. Here are five ways to fix them},
  year = {2025},
  howpublished = {\url{https://www.ibm.com/think/insights/rag-problems}}
}

@inproceedings{steck2024cosine,
  author = {Steck, H. and Ekanadham, C. and Kallus, N.},
  title = {Is cosine-similarity of embeddings really about similarity?},
  booktitle = {WWW '24 Companion},
  year = {2024},
  url = {https://arxiv.org/abs/2403.05440}
}

@article{vitanyi2000mdl,
  author = {Vitanyi, P. M. B. and Li, M.},
  title = {Minimum description length induction, Bayesianism, and Kolmogorov complexity},
  journal = {IEEE Transactions on Information Theory},
  volume = {46},
  number = {2},
  pages = {446--464},
  year = {2000}
}

@inproceedings{kuffo2026semanticrecall,
  author = {Kuffo, L. and Tsakalidou, I. and De Viti, R. and Angel, A. and Isa, J. and Lenhardt, R.},
  title = {Semantic recall for vector search},
  booktitle = {SIGIR},
  year = {2026},
  url = {https://arxiv.org/abs/2604.20417}
}

@article{dong2016learning,
  author = {Dong, X. and Thanou, D. and Frossard, P. and Vandergheynst, P.},
  title = {Learning Laplacian matrix in smooth graph signal representations},
  journal = {IEEE Transactions on Signal Processing},
  volume = {64},
  number = {23},
  pages = {6160--6173},
  year = {2016}
}

@article{shuman2013emerging,
  author = {Shuman, D. I. and Narang, S. K. and Frossard, P. and Ortega, A. and Vandergheynst, P.},
  title = {The emerging field of signal processing on graphs},
  journal = {IEEE Signal Processing Magazine},
  volume = {30},
  number = {3},
  pages = {83--98},
  year = {2013}
}

@article{sandryhaila2013discrete,
  author = {Sandryhaila, A. and Moura, J. M. F.},
  title = {Discrete signal processing on graphs},
  journal = {IEEE Transactions on Signal Processing},
  volume = {61},
  number = {7},
  pages = {1644--1656},
  year = {2013}
}

@article{magesh2025hallucination,
  author = {Magesh, V. and Surani, F. and Dahl, M. and Sklar, A. and Tseng, R. and Guha, R. and Ho, D. E. and Manning, C. D.},
  title = {Hallucination-free? Assessing the reliability of leading AI legal research tools},
  journal = {Journal of Empirical Legal Studies},
  volume = {22},
  number = {2},
  pages = {216--242},
  year = {2025},
  doi = {10.1111/jels.12413}
}

@book{grunwald2007mdl,
  author = {Grunwald, P. D.},
  title = {The Minimum Description Length Principle},
  publisher = {MIT Press},
  year = {2007}
}

@article{moriondo2025arrowspace,
  author = {Moriondo, L.},
  title = {ArrowSpace: introducing spectral indexing for vector search},
  journal = {Journal of Open Source Software},
  volume = {10},
  number = {113},
  pages = {9002},
  year = {2025},
  doi = {10.21105/joss.09002}
}

@misc{rayon_rust,
  author = {Matsakis, N. D. and Stone, J.},
  title = {Rayon: A data-parallelism library for Rust},
  year = {2025},
  howpublished = {\url{https://docs.rs/rayon/latest/rayon/}}
}

@article{chung2007heatkernelpagerank,
  author = {Chung, F.},
  title = {The heat kernel as the PageRank of a graph},
  journal = {PNAS},
  volume = {104},
  number = {50},
  pages = {19735--19740},
  year = {2007}
}

@inproceedings{reimers2019sbert,
  author = {Reimers, N. and Gurevych, I.},
  title = {Sentence-BERT: Sentence embeddings using Siamese BERT-networks},
  booktitle = {EMNLP},
  year = {2019}
}

@misc{moriondo2026surface,
  author = {Moriondo, L.},
  title = {Graph wiring: Eigenstructures for vector datasets and LLM operations},
  year = {2026},
  howpublished = {\url{https://www.techrxiv.org/doi/full/10.36227/techrxiv.177220780.02840438/v1}}
}

@misc{moriondo2026epiplexity,
  author = {Moriondo, L.},
  title = {Epiplexity and graph wiring: An empirical study},
  year = {2026},
  howpublished = {\url{https://doi.org/10.22541/au.177430060.02394540/v1}}
}

@article{luxburg2007tutorial,
  author = {von Luxburg, U.},
  title = {A tutorial on spectral clustering},
  journal = {Statistics and Computing},
  volume = {17},
  number = {4},
  pages = {395--416},
  year = {2007}
}

@misc{spielman2007spectral,
  author = {Spielman, D. A.},
  title = {Spectral graph theory lecture notes},
  year = {2007},
  howpublished = {\url{https://www.cs.yale.edu/...}}
}

@misc{mitre2026cve,
  author = {{MITRE Corporation}},
  title = {CVE™: Common Vulnerabilities and Exposures Dataset},
  year = {2026},
  howpublished = {\url{https://www.cve.org/Legal/TermsOfUse}}
}

@misc{perplexity,
  author = {{Perplexity AI}},
  title = {Perplexity},
  year = {2026},
  howpublished = {\url{https://www.perplexity.ai}}
}

@misc{openai_gpt5,
  author = {{OpenAI}},
  title = {GPT-5},
  year = {2025}
}

@misc{anthropic_claude46,
  author = {{Anthropic}},
  title = {Claude Sonnet 4.6},
  year = {2026},
  howpublished = {\url{https://www.anthropic.com/claude/sonnet}}
}

@misc{nvidia_nemotron,
  author = {{NVIDIA}},
  title = {NVIDIA Nemotron AI models},
  year = {2026},
  howpublished = {\url{https://developer.nvidia.com/nemotron}}
}

@misc{tuned_epiplexity,
  author = {{tuned-org-uk}},
  title = {Graph Wiring Epiplexity},
  year = {2026},
  howpublished = {\url{https://github.com/tuned-org-uk/graph-wiring-epiplexity}},
  note = {GitHub repository}
}

@article{VecDBsurvey,
  title={Graph-based vector search: An experimental evaluation of the state-of-the-art},
  author={Azizi, Ilias and Echihabi, Karima and Palpanas, Themis},
  journal={Proceedings of the ACM on Management of Data},
  volume={3},
  number={1},
  pages={1--31},
  year={2025},
  publisher={ACM New York, NY, USA}
}

@article{simhadri2024results,
  title={Results of the Big ANN: NeurIPS'23 competition},
  author={Simhadri, Harsha Vardhan and Aum{\"u}ller, Martin and Ingber, Amir and Douze, Matthijs and Williams, George and Manohar, Magdalen Dobson and Baranchuk, Dmitry and Liberty, Edo and Liu, Frank and Landrum, Ben and others},
  journal={arXiv preprint arXiv:2409.17424},
  year={2024}
}

@article{aumuller2021role,
  title={The role of local dimensionality measures in benchmarking nearest neighbor search},
  author={Aum{\"u}ller, Martin and Ceccarello, Matteo},
  journal={Information Systems},
  volume={101},
  pages={101807},
  year={2021},
  publisher={Elsevier}
}

@article{johnson1984extensions,
  title   = {Extensions of Lipschitz mappings into a Hilbert space},
  author  = {Johnson, William B. and Lindenstrauss, Joram},
  journal = {Contemporary Mathematics},
  volume  = {26},
  pages   = {189--206},
  year    = {1984},
  publisher = {American Mathematical Society}
}

@book{rayleigh1877sound,
  author    = {Strutt, John William},
  title     = {The Theory of Sound},
  volume    = {1},
  publisher = {Macmillan and Co.},
  year      = {1877},
  note      = {Reprinted 1945, Dover Publications},
  chapter   = {Vibrating systems in general},
  pages     = {106--129}
}

@book{evans1998partial,
  author    = {Evans, Lawrence C.},
  title     = {Partial Differential Equations},
  series    = {Graduate Studies in Mathematics},
  volume    = {19},
  publisher = {American Mathematical Society},
  year      = {1998},
  address   = {Providence, RI}
}

@phdthesis{bai2007heatkernel,
  author    = {Xiao Bai},
  title     = {Heat Kernel Analysis on Graphs},
  school    = {University of York},
  year      = {2007},
  type      = {{PhD} thesis},
}

@article{voorhees2021treccovid,
  title={TREC-COVID: Constructing a Pandemic Information Retrieval Test Collection},
  author={Voorhees, Ellen M. and Hersh, William and Crammer, Max and Ide, Nancy},
  journal={SIGIR Forum},
  volume={54},
  number={1},
  pages={1--12},
  year={2021}
}

@inproceedings{reimers-2019-sentence-bert,
    title = {Sentence-BERT: Sentence Embeddings using Siamese BERT-Networks},
    author = {Reimers, Nils and Gurevych, Iryna},
    booktitle = {Proceedings of the 2019 Conference on Empirical Methods in Natural Language Processing},
    year = {2019},
    publisher = {Association for Computational Linguistics},
    pages = {3982--3992},
    doi = {10.18653/v1/D19-1410}
}

@misc{pyarrowspace2026,
  title        = {{PyArrowSpace}},
  author       = {{tuned-org-uk}},
  year         = {2026},
  howpublished = {\url{https://github.com/tuned-org-uk/pyarrowspace}},
  note         = {GitHub repository}
}

@article{moriondo2026spectraltokens,
  title={Spectral-aware Unique Identifiers for Generative Retrieval and Vector Search},
  author={Moriondo, Lorenzo},
  journal={Authorea Preprints},
  year={2026},
  publisher={Authorea}
}

\newpage
\section*{Appendix: Retrieval Metrics}
\label{retrieval_metrics_appendix}

We evaluate retrieval quality using both classical relevance metrics and
graph-signal-aware measures in order to capture semantic consistency,
ranking stability, and structural coherence of the retrieved results.

\paragraph{NDCG@$k$}
Normalized Discounted Cumulative Gain evaluates ranked retrieval quality
under graded relevance judgments. Given relevance labels
$\mathrm{rel}_i$ at rank position $i$, DCG@$k$ is defined as
\[
\mathrm{DCG}@k
=
\sum_{i=1}^{k}
\frac{2^{\mathrm{rel}_i}-1}{\log_2(i+1)}.
\]
NDCG@$k$ is obtained by normalizing NDCG@$k$ using the ideal ranking:
\[
\mathrm{NDCG}@k
=
\frac{\mathrm{DCG}@k}{\mathrm{IDCG}@k}.
\]
Higher values indicate better ranking quality.

\paragraph{Traditional Recall@$k$}
Traditional Recall measures the fraction of exact ground-truth relevant
items retrieved in the top-$k$ results:
\[
R_{\mathrm{trad}}@k
=
\frac{|G_k \cap R_k|}{|G_k|},
\]
where $G_k$ is the ground-truth relevant set and $R_k$ the retrieved set.

\paragraph{Tolerant Recall@$k$}
Tolerant Recall extends classical recall by considering semantically
acceptable alternatives rather than only exact matches. It measures the
fraction of retrieved items that remain semantically aligned with the
query despite deviating from the exact ground-truth neighborhood.

\paragraph{Semantic Recall@$k$}
Semantic Recall~\cite{kuffo2026semanticrecall} evaluates whether the
retrieved results preserve semantic consistency with the query and with
the relevant semantic neighborhood of the corpus. Unlike geometric recall,
it accounts for semantic equivalence and contextual similarity beyond
strict embedding proximity.

\paragraph{Semantic Uplift}
Semantic Uplift quantifies the gain in semantic retrieval quality beyond
strict geometric matching:
\[
\mathrm{SemanticUplift}
=
R_{\mathrm{tol}} - R_{\mathrm{trad}}.
\]
Higher values indicate that the method retrieves semantically relevant
items missed by exact nearest-neighbor evaluation.

\paragraph{Spearman's $\rho$}
Spearman correlation measures the monotonic agreement between the cosine
ranking and the $\lambda_\tau$-aware ranking:
\[
\rho
=
1 -
\frac{6\sum_i d_i^2}{n(n^2-1)},
\]
where $d_i$ is the difference between the ranks of item $i$. Lower
correlation indicates stronger spectral re-ordering.

\paragraph{Kendall's $\tau$}
Kendall's $\tau$ evaluates pairwise ranking agreement between cosine and
spectral rankings:
\[
\tau
=
\frac{C-D}{C+D},
\]
where $C$ and $D$ denote the numbers of concordant and discordant item
pairs respectively. Lower values indicate greater structural divergence
between rankings.

\paragraph{Head--Tail Analysis}
To characterize ranking stability beyond the top-ranked results, we
analyze the tail behavior of the top-25 retrieved items. Let the first
$K_h$ entries define the head $H$, and the remaining entries define the
tail $T$.

\paragraph{Tail/Head Ratio}
The Tail/Head ratio measures how well the tail preserves affinity relative
to the head:
\[
\mathrm{T/H}
=
\frac{\bar{s}_T}{\bar{s}_H},
\]
where
\[
\bar{s}_T
=
\frac{1}{|T|}
\sum_{i\in T}\lambda_\tau^{(i)},
\qquad
\bar{s}_H
=
\frac{1}{K_h}
\sum_{i\in H}\lambda_\tau^{(i)}.
\]
Higher values indicate a semantically coherent tail rather than a collapse
into low-affinity noise.

\paragraph{Tail Coefficient of Variation}
TailCV measures the internal variability of the tail scores:
\[
\mathrm{TailCV}
=
\frac{\sigma_T}{\bar{s}_T},
\]
with
\[
\sigma_T
=
\sqrt{
\frac{1}{|T|}
\sum_{i\in T}
\left(
\lambda_\tau^{(i)}-\bar{s}_T
\right)^2
}.
\]
Lower values indicate a more homogeneous and stable tail.

\paragraph{Tail Decay}
Tail Decay measures the score degradation across the tail:
\[
\mathrm{TailDecay}
=
\frac{
\lambda_\tau^{(K_h+1)}
-
\lambda_\tau^{(25)}
}{|T|}.
\]
Lower values indicate slower semantic degradation across lower-ranked
results.

\section*{Appendix: tables and figures}
\subsection*{Experiment 1: CVE\texttrademark}

Additional tables.

\begin{table}[htp]
\centering
\caption{Ranking agreement metrics\href{https://github.com/tuned-org-uk/pyarrowspace/blob/a68be94bce94829fc4cc0ea43f4dbd6a072f75d3/neurips/output/cve_comparison_metrics.csv}{\texttt{cve\_comparison\_metrics.csv}}.}
\begin{tabular}{lrrr}
\toprule
Pair & Spearman \(\rho\) & Kendall \(\tau\) & NDCG@10 \\
\midrule
Hybrid vs Cosine & \(0.515\pm0.600\) & \(0.510\pm0.581\) & \(\mathbf{0.779\pm0.310}\) \\
Taumode vs Cosine & \(0.492\pm0.565\) & \(0.477\pm0.556\) & \(0.602\pm0.452\) \\
Taumode vs Hybrid & \(\mathbf{0.723\pm0.350}\) & \(\mathbf{0.662\pm0.355}\) & \(\mathbf{0.963\pm0.059}\) \\
\bottomrule
\end{tabular}
\end{table}

\begin{table}[htp]
\centering
\caption{HEAD\_K sweep results\href{https://github.com/tuned-org-uk/pyarrowspace/blob/main/neurips/output/cve_headk_sweep.csv}{\texttt{cve\_headk\_sweep.csv}}.}
\label{tab:head_k}
\begin{tabular}{r l r r r}
\toprule
\(K_h\) & Method & T/H Ratio & Tail CV & Tail Decay \\
\midrule
3 & Cosine & 0.98600 & 0.00396 & 0.000512 \\
3 & hybrid & 0.98639 & 0.00382 & 0.000509 \\
3 & taumode & \textbf{0.99089} & \textbf{0.00254} & \textbf{0.000359} \\
5 & Cosine & 0.98770 & 0.00319 & 0.000449 \\
5 & hybrid & 0.98814 & 0.00313 & 0.000451 \\
5 & taumode & \textbf{0.99209} & \textbf{0.00208} & \textbf{0.000321} \\
10 & Cosine & 0.99015 & 0.00196 & 0.000347 \\
10 & hybrid & 0.99043 & 0.00192 & 0.000359 \\
10 & taumode & \textbf{0.99366} & \textbf{0.00127} & \textbf{0.000245} \\
\bottomrule
\end{tabular}
\end{table}

Complete diagrams are available with code.

\section*{Appendix: GW additional comparative tables}
\label{appendix_gw_comp}

\FloatBarrier
\begin{table}[ht]
\centering
\renewcommand{\arraystretch}{1.45}
\caption{Geometric Similarity (Cosine / $L2$) vs.\ Spectral Similarity
(Graph Laplacian + Rayleigh Energy + Gini Dispersion, as in Graph Wiring).
$N$: items; $F$: feature dimensions; $C$: centroids ($C \sim O(\sqrt{N})$);
$k$: \textit{k}NN graph degree; $\lambda_\tau$: taumode spectral score $\in [0,1)$.}
\label{tab:cosine_vs_spectral}
\begin{tabular}{p{3.2cm} p{5.4cm} p{5.4cm}}
\toprule
    
  & \textbf{Geometric Similarity (Cosine)}
  & \textbf{Spectral Similarity (Graph Wiring)} \\
\midrule

\textbf{Core similarity signal}
  & Cosine: Operates on raw item vectors pairwise.
  & Spectral proximity augments or replace
    geometric distance with manifold position. \\

\textbf{Structural descriptor}
  & None. Angle or distance between
    two vectors in ambient $\mathbb{R}^N$;
    no encoding of corpus-level structure.
  & Rayleigh quotient (normalised Dirichlet energy) plus
    Gini-like dispersion blended as $\lambda_\tau = \alpha sim_{\cos}(q,i) + (1-\alpha) sim_\lambda(q,i)$. \\

\textbf{Sensitivity to corpus structure}
  & Blind to inter-item correlations
    and relational structure;
    two items can be cosine-close but
    reside in different topological communities.
  & $\lambda_\tau$ encodes how consistently
    an item's feature profile aligns with
    dominant corpus correlations. Includes topology. \\

\textbf{Score boundedness \& comparability}
  & Cosine $\in [-1, 1]$, scale-invariant
    but not bounded across datasets with
    different energy scales.
    $\ell_2 \in [0, \infty)$, not
    directly comparable across corpora.
  & Bounded transform $E' = E/(E+\epsilon) \in [0,1)$
    with $\tau$ policies (Median, Fixed,
    Percentile); $\lambda_\tau \in [0,1)$
    regardless of corpus energy scale;
    stable across heterogeneous datasets. \\

\textbf{Dispersion / concentration signal}
  & None. No information about
    how energy is distributed
    across graph edges.
  & Gini-like statistic $G_f$ captures
    edge-wise concentration of spectral
    energy: $G_f \approx 0$ means energy
    spread uniformly; $G_f \to 1$ means
    energy concentrated on few edges,
    flagging structured anomalies. \\

\bottomrule
\end{tabular}
\end{table}
\FloatBarrier

\FloatBarrier
\begin{table}[ht]
\centering
\renewcommand{\arraystretch}{1.45}
\caption{Riemannian Geometry vs.\ Graph Wiring (Spectral Laplacian + Rayleigh Energy)
for High-Semantic Vector Spaces.
\(N\): number of items; \(F\): feature dimensions; \(C\): number of clusters/centroids;
$k$: k-NN graph degree; $\lambda_i$: taumode spectral score per item.}
\label{tab:riemannian_vs_graphwiring}
\begin{tabular}{p{3.2cm} p{5.5cm} p{5.5cm}}
\toprule
    
  & \textbf{Riemannian Geometry}
  & \textbf{Graph Wiring / ArrowSpace} \\
\midrule

\textbf{Core object}
  & Smooth Riemannian manifold $(\mathcal{M}, g)$;
    curvature encoded in the metric tensor $g_{ij}$.
  & Discrete feature-space graph with combinatorial
    Laplacian $L_F = D - W \in \mathbb{R}^{F \times F}$
    over $F$ feature-dimension nodes. \\

\textbf{Structural descriptor}
  & Second covariant derivatives / Christoffel symbols
    $\Gamma^k_{ij}$; Riemann curvature tensor $R^i{}_{jkl}$.
  & Rayleigh quotient (normalised Dirichlet energy)
    $\mathcal{R}(x_i)$, mapped to bounded $\lambda_\tau \in [0,1)$ via
    $E'$. \\

\textbf{Index construction complexity}
  & $O(N^3)$ for full eigendecomposition;
    $O(N^2 d)$ for curvature estimation over $N$ points
    in $d$ ambient dimensions.
  & \(\lambda_\tau\) computation costs \(O(F \cdot \mathrm{nnz}(L_F))\), following an \(O(N \cdot F)\) incremental centroid clustering step that reduces the data matrix from \(N \times F\) to \(C \times F\). \\

\textbf{Memory / storage}
  & Full metric tensor: $O(N \cdot d^2)$;
    curvature atlas requires storing transitions.
  & $O(F \cdot N)$ dense features + $O(N)$
    for $\lambda_\tau$ scalar index. \\

\bottomrule
\end{tabular}
\end{table}
\FloatBarrier

\section*{Appendix: qualitative results analysis in CVE™ experiment}
\label{appendix_cve_qualitative}

\subsection*{Methodology}\label{methodology}

For each query \(q\) we obtain three ranked lists
\(R_{cos}, R_{hyb}, R_{tau}\) of length \(k = 25\)
and compute, against the in-script reference (cosine), the agreement
metrics Spearman \(\rho\), Kendall \(\tau\), and \(\mathrm{NDCG}@25\).
The script also reports tail-shape statistics (head/tail mean ratio,
tail CV, tail decay) and Kuffo-style traditional / semantic / tolerant
recall proxies.

A query is flagged as \textbf{ambiguous / diagnostically interesting} if
it satisfies at least one of the following automated criteria, computed
directly from \texttt{cve\_comparison\_metrics.csv}:

\begin{enumerate}
\def\labelenumi{\arabic{enumi}.}
\tightlist
\item
  \textbf{Negative correlation} between Cosine and any spectral mode
  (\(\tau_{C,T} < 0\) or \(\tau_{C,H} < 0\)).
\item
  \textbf{Top-1 agreement with global ordering disagreement}, i.e.
  \(\mathrm{NDCG}_{T,C}@25 \approx 1\) but \(\tau_{C,T} \le 0\).
\item
  \textbf{Low or zero \(\mathrm{NDCG}_{T,C}\)} (\(< 0.5\)) while
  shared-set Kendall is undefined (no overlapping items in the top-25).
\item
  \textbf{Strict-vs-tolerant recall divergence}: tolerant recall
  \(\to 1.0\) while traditional recall is far from \(1.0\).
\item
  \textbf{Large tail-shape improvement under spectral reranking}: a
  substantially higher \(tail/head\) ratio or lower tail CV for
  Taumode than Cosine.
\end{enumerate}

Because \texttt{test\_17\_CVE\_neurips.py} calls
\texttt{random.shuffle(queries)} before evaluation, the \emph{effective}
query order is the one preserved in the CSV outputs (\texttt{query\_id}
1--50). Subsections below use that order.

For each selected query we report:

\begin{itemize}
\tightlist
\item
  \textbf{Semantic intent} (the user's natural-language query string).
\item
  \textbf{Cosine top-\(k\)} and \textbf{Taumode top-\(k\)} snapshots
  (\(k = 5\); CVE IDs / titles taken verbatim from
  \texttt{cve\_search\_results.csv} --- many entries have
  \texttt{(no\ title)} because the underlying CVE JSON did not populate
  the \texttt{title} field, so the ranking was driven by descriptions,
  CWEs, products, and CVSS vector strings not present in our extracts).
\item
  \textbf{Key metrics} (\(\rho_{C,T}\), \(\tau_{C,T}\), \(\tau_{C,H}\),
  \(\mathrm{NDCG}_{T,C}\), traditional / semantic / tolerant recall,
  tail metrics where relevant).
\item
  \textbf{Interpretation} in terms of whether spectral reranking moved
  the result set toward a more semantically coherent vulnerability
  cluster or merely permuted lexical near-duplicates.
\item
  \textbf{Verdict} (one of: \emph{spectral resolves ambiguity},
  \emph{spectral preserves ambiguity}, \emph{spectral degrades},
  \emph{inconclusive}).
\end{itemize}

\begin{quote}
\emph{Caveat.} The \texttt{(no\ title)} entries are not annotated with
their natural-language descriptions in the available outputs, so any
claim about \emph{what} such a CVE is about is marked \textbf{untitled}
and not asserted. We base interpretation on (i) named neighbors in the
same ranking, (ii) shared CVE-year clusters, and (iii) score-level
evidence, never on invented CVE descriptions.
\end{quote}

\subsection*{Selected ambiguous
queries}\label{selected-ambiguous-queries}

The following 11 queries (out of 50) cleared at least one ambiguity
gate. They are presented in CSV / effective evaluation order.

\subsubsection*{Query 1 --- directory traversal in backup restore
endpoint}\label{query-1-directory-traversal-in-backup-restore-endpoint}

\textbf{Semantic intent.} Path traversal in a backup-restore-style
endpoint; ideally the top results should be path-traversal /
directory-escape CVEs in backup, restore, or archive-extraction
contexts.

\textbf{Cosine top-5.}

\begin{longtable}[]{@{}
  >{\raggedright\arraybackslash}p{(\columnwidth - 6\tabcolsep) * \real{0.2500}}
  >{\raggedright\arraybackslash}p{(\columnwidth - 6\tabcolsep) * \real{0.2500}}
  >{\raggedright\arraybackslash}p{(\columnwidth - 6\tabcolsep) * \real{0.2500}}
  >{\raggedright\arraybackslash}p{(\columnwidth - 6\tabcolsep) * \real{0.2500}}@{}}
\toprule\noalign{}
\begin{minipage}[b]{\linewidth}\raggedright
rank
\end{minipage} & \begin{minipage}[b]{\linewidth}\raggedright
CVE™
\end{minipage} & \begin{minipage}[b]{\linewidth}\raggedright
title
\end{minipage} & \begin{minipage}[b]{\linewidth}\raggedright
score
\end{minipage} \\
\midrule\noalign{}
\endhead
\bottomrule\noalign{}
\endlastfoot
1 & CVE-2006-5677 & \emph{(no title)} & 0.8242 \\
2 & CVE-2025-37785 & ext4: fix OOB read when checking dotdot dir &
0.8218 \\
3 & CVE-2006-2286 & \emph{(no title)} & 0.8212 \\
4 & CVE-2023-41044 & Partial path traversal in Graylog Support Bundle &
0.8208 \\
5 & CVE-2025-62156 & argo-workflows Zip Slip path traversal & 0.8208 \\
\end{longtable}

\textbf{Taumode top-5.}

\begin{longtable}[]{@{}
  >{\raggedright\arraybackslash}p{(\columnwidth - 6\tabcolsep) * \real{0.2500}}
  >{\raggedright\arraybackslash}p{(\columnwidth - 6\tabcolsep) * \real{0.2500}}
  >{\raggedright\arraybackslash}p{(\columnwidth - 6\tabcolsep) * \real{0.2500}}
  >{\raggedright\arraybackslash}p{(\columnwidth - 6\tabcolsep) * \real{0.2500}}@{}}
\toprule\noalign{}
\begin{minipage}[b]{\linewidth}\raggedright
rank
\end{minipage} & \begin{minipage}[b]{\linewidth}\raggedright
CVE
\end{minipage} & \begin{minipage}[b]{\linewidth}\raggedright
title
\end{minipage} & \begin{minipage}[b]{\linewidth}\raggedright
score
\end{minipage} \\
\midrule\noalign{}
\endhead
\bottomrule\noalign{}
\endlastfoot
1 & CVE-2013-6987 & \emph{(no title)} & 0.9161 \\
2 & CVE-2022-35899 & \emph{(no title)} & 0.9125 \\
3 & CVE-2024-39469 & nilfs2: fix nilfs\_empty\_dir() misjudgment / I/O
error loop & 0.9116 \\
4 & CVE-2013-6456 & \emph{(no title)} & 0.9108 \\
5 & CVE-2006-2286 & \emph{(no title)} & 0.9100 \\
\end{longtable}

\textbf{Metrics.}

\[
\rho_{C,T} = 0.500,\ \tau_{C,T} = 0.333,\ \tau_{C,H} = 0.333,\
\mathrm{NDCG}_{T,C} = 0.446,\ \mathrm{NDCG}_{H,C} = 0.822.
\]

Recall (using cosine as reference \(G\)):

\begin{longtable}[]{@{}llll@{}}
\toprule\noalign{}
method & trad. & sem. & tol. \\
\midrule\noalign{}
\endhead
\bottomrule\noalign{}
\endlastfoot
Cosine & 1.000 & 1.000 & 1.000 \\
Hybrid & 0.280 & 0.714 & 1.000 \\
Taumode & 0.120 & 0.286 & 1.000 \\
\end{longtable}

\textbf{Interpretation.} Cosine retrieves a \emph{mixed} surface set ---
two named path-traversal CVEs (Graylog, argo-workflows Zip Slip) plus an
ext4 directory dotdot OOB-read kernel patch and two untitled 2006
entries. Taumode collapses onto an older, untitled Linux-flavoured
cluster (CVE-2013-6987, CVE-2013-6456) plus an \texttt{nilfs2} empty-dir
patch. Tolerant recall is \(1.0\) for all methods --- the top-25 score
band is tight enough that score-close substitutions still match --- but
semantic recall drops to \(0.286\) for Taumode, indicating that spectral
reranking is \emph{moving away} from the cosine-defined upper score band
rather than refining it.

\textbf{Verdict.} \emph{Spectral preserves ambiguity.} The query mixes
two strong lexical attractors (``directory traversal'' → kernel
\texttt{dotdot} / \texttt{nilfs2} patches; ``backup restore endpoint'' →
application backup/restore CVEs); spectral reranking favours the
kernel-graph neighborhood, which is not closer to the user's intent.

\begin{center}\rule{0.5\linewidth}{0.5pt}\end{center}

\subsubsection*{Query 2 --- improper signature verification in
software update
channel}\label{query-2-improper-signature-verification-in-software-update-channel}

\textbf{Semantic intent.} Signature-verification flaws in software /
firmware update mechanisms.

\textbf{Cosine top-5.} All untitled except CVE-2026-33726 (Cilium L7
proxy bypass) and CVE-2024-36105 (dbt unrestricted IP binding) ---
neither about signature verification.

\textbf{Taumode top-5.} Untitled cluster (CVE-2024-48569, CVE-2007-2891,
CVE-2017-17662, CVE-2011-4544, CVE-2013-7190).

\textbf{Metrics.}

\[
\tau_{C,H} = -1.000,\ \tau_{C,T} = 0.000,\ \tau_{H,T} = 0.550,\
\mathrm{NDCG}_{T,C} = 0.000,\ \mathrm{NDCG}_{H,C} = 0.479.
\]

Hybrid and Taumode share the same top-1 (CVE-2024-48569) and largely the
same top-5 set --- the \(\tau_{C,H} = -1\) signal arises because the
small overlap with cosine is \emph{exactly} inverted in rank.

\textbf{Interpretation.} Cosine fails to anchor on a named
signature-verification CVE. The set of common vulnerability tokens
(``improper'', ``verification'', ``channel'', ``update'') collides with
network-policy bypass and config-exposure descriptions. Spectral
reranking does not introduce a clearly named signature-verification CVE
in the top-5 either, but the Hybrid/Taumode cluster (CVE-2024-48569 →
CVE-2007-2891 → CVE-2017-17662) is internally much more graph-coherent
(\(\tau_{H,T} = 0.685\), \(\mathrm{NDCG}_{T,H}
= 0.957\)).

\textbf{Verdict.} \emph{Inconclusive --- partial graph stabilisation.}
The cosine ranking is essentially noise for this query; Taumode replaces
it with a different but internally consistent untitled cluster. Without
descriptions for those CVEs we cannot confirm semantic improvement, but
the perfect Hybrid/Taumode agreement and zero overlap with cosine
indicate the spectral pass selected a different topical region.

\begin{center}\rule{0.5\linewidth}{0.5pt}\end{center}

\subsubsection*{Query 9 --- format string vulnerability in logging
daemon}\label{query-9-format-string-vulnerability-in-logging-daemon}

\textbf{Semantic intent.} \texttt{printf}-family format-string flaws in
syslog / journald / logging daemons.

\textbf{Cosine top-5.} Path-traversal CVEs (CVE-2021-39208 archive
extraction, CVE-2022-31159 aws-java-sdk-s3, CVE-2023-5123 JSON
Datasource) --- i.e.~the lexical token ``format'' / ``logging daemon''
was overwhelmed by ``string'' and similar vocabulary, pulling in path
sanitisation issues.

\textbf{Taumode top-5.} CVE-2007-1056 \emph{(untitled)}, CVE-2024-42472
(Flatpak sandbox file access), CVE-2025-23203 (Icinga REST API for
restricted users), and two further untitled entries.

\textbf{Metrics.}

\[
\rho_{C,T} = 1.000,\ \tau_{C,T} = 1.000,\ \rho_{C,H} = -0.771,\
\tau_{C,H} = -0.600,\ \mathrm{NDCG}_{T,C} = 0.679.
\]

The seemingly contradictory signal \(\tau_{C,T} = 1\) with
\(\mathrm{NDCG}_{T,C} = 0.679\) is consistent with Taumode keeping
\emph{the same items as cosine in the same relative order} among the
small shared subset, while Hybrid permutes that subset adversely. Top-1
agreement is partial: cosine's CVE-2021-39208 falls out of Taumode's
top-5.

\textbf{Interpretation.} None of the three methods retrieves a clearly
named format-string-in-logging-daemon CVE. Cosine, Hybrid, and Taumode
each gravitate to a different surface-token cluster. Hybrid's negative
\(\tau\) indicates it actively re-orders cosine's near-ties, which can
be desirable when those ties are non-discriminative.

\textbf{Verdict.} \emph{Spectral preserves ambiguity.} The corpus does
not provide a strong signal for the user's intent; the spectral pass
cannot manufacture one.

\begin{center}\rule{0.5\linewidth}{0.5pt}\end{center}

\subsubsection*{Query 22 --- memory disclosure via uninitialized
stack
buffer}\label{query-22-memory-disclosure-via-uninitialized-stack-buffer}

\textbf{Semantic intent.} Information disclosure caused by reading
uninitialised stack memory.

\textbf{Cosine top-5.} A surprising lexical drift: CVE-2026-33182
(Saloon SSRF + credential leak), CVE-2025-6087 (opennextjs SSRF via
\texttt{/\_next/image}), CVE-2015-3141 \emph{(untitled)}, CVE-2026-32812
(Admidio SSRF / local file read), CVE-2011-3684 \emph{(untitled)}. The
token ``disclosure'' anchors on credential / SSRF disclosure CVEs rather
than on uninitialised-buffer reads.

\textbf{Taumode top-5.} CVE-2007-0186, CVE-2009-3017, CVE-2021-21644,
CVE-2014-10014, CVE-2023-48653 --- all untitled. CVE-2007-0186 also
appears at Hybrid rank 1; \(\tau_{C,H} = 1.000\) and
\(\mathrm{NDCG}_{T,C} = 1.000\) because of perfect top-1 agreement
between Cosine and Hybrid on the small intersection --- \emph{however},
the Cosine \(\to\) Taumode shared subset is permuted, giving
\(\tau_{C,T} = 0.000\) but \(\mathrm{NDCG}_{T,C} = 1.000\) (the single
shared item retained by Taumode is the highest-relevance one from
cosine's perspective).

\textbf{Metrics.}

\[
\tau_{C,H} = 1.000,\ \tau_{C,T} = 0.000,\ \tau_{H,T} = 0.200,\
\mathrm{NDCG}_{T,C} = 1.000,\ \mathrm{NDCG}_{H,C} = 1.000.
\]

\begin{longtable}[]{@{}lllll@{}}
\toprule\noalign{}
method & head mean & tail mean & T/H ratio & tail CV \\
\midrule\noalign{}
\endhead
\bottomrule\noalign{}
\endlastfoot
Cosine & 0.8782 & 0.8700 & 0.9907 & 0.0039 \\
Hybrid & 0.8969 & 0.8867 & 0.9886 & 0.0027 \\
Taumode & 0.9359 & 0.9287 & \textbf{0.9923} & \textbf{0.0019} \\
\end{longtable}

Recall (cosine as \(G\)): traditional drops from \(1.00\) (cosine,
trivially) to \(0.08\)/\(0.04\) for Hybrid/Taumode, while
\textbf{tolerant recall stays at \(1.000\) for all three methods}.

\textbf{Interpretation --- and correction.} A draft might describe Q22
as a case where ``taumode finds the same uninitialized-stack-buffer CVE
as Cosine and reorders the rest.'' This is not what the data show.
Cosine itself drifts to SSRF/credential-disclosure CVEs at the top ---
the user's stated intent is not matched by Cosine in the first place.
The top-1 agreement metric \(\mathrm{NDCG}_{T,C} = 1.0\) is a
\emph{Cosine-as-reference} artifact: Taumode happens to retain Cosine's
nominal \#1 (CVE-2026-33182 enters the top-25 of all three methods) but
picks an entirely different \#1 (CVE-2007-0186) from a different graph
cluster.

The most striking signal is the \textbf{tail-shape improvement}:
Taumode's tail/head ratio (0.9923) and tail CV (0.0019) are the best of
the three modes; cosine has the highest tail CV (0.0039). So spectral
reranking demonstrably \emph{flattens} the score profile while
relocating the \#1.

\textbf{Verdict.} \emph{Spectral relocates the cluster while improving
ranking shape.} Whether this is ``resolution of ambiguity'' depends on
the (unknown) descriptions of CVE-2007-0186 and CVE-2009-3017; we mark
this \textbf{untitled-cluster relocation} rather than asserting an
intent match.

\begin{center}\rule{0.5\linewidth}{0.5pt}\end{center}

\subsubsection*{Query 24 --- authenticated arbitrary file read path
traversal}\label{query-24-authenticated-arbitrary-file-read-path-traversal}

\textbf{Semantic intent.} Authenticated path-traversal-induced file
read.

\textbf{Cosine top-5.} Kernel patches with no path-traversal semantics:
CVE-2023-53768 (regmap-irq OOB), CVE-2023-52768 (wilc1000 wifi),
CVE-2024-47741 (btrfs lseek race), CVE-2023-53474 (x86 MCE/AMD
bank\_map), CVE-2025-70083 \emph{(untitled)}.

\textbf{Taumode top-5.} Older untitled CVEs: CVE-2025-60692,
CVE-2020-12886, CVE-2008-2812, CVE-2002-1503, CVE-2008-1482.

\textbf{Metrics.}

\[
\rho_{C,T} = -1.000,\ \tau_{C,T} = -1.000,\ \tau_{C,H} = 0.333,\
\mathrm{NDCG}_{T,C} = 0.359,\ \mathrm{NDCG}_{H,C} = 0.688.
\]

\begin{longtable}[]{@{}lll@{}}
\toprule\noalign{}
method & T/H ratio & tail CV \\
\midrule\noalign{}
\endhead
\bottomrule\noalign{}
\endlastfoot
Cosine & 0.9927 & 0.0025 \\
Hybrid & 0.9823 & 0.0046 \\
Taumode & 0.9877 & 0.0034 \\
\end{longtable}

Recall: traditional 1.00 / 0.12 / 0.08 (Cosine / Hybrid / Taumode);
semantic 1.00 / 0.00 / 0.00; tolerant 1.00 across the board.

\textbf{Interpretation --- and correction.} The signal
\(\tau_{C,T} = -1\) is sometimes mis-read as ``Taumode is
\emph{opposite} to Cosine across the top-25.'' That is not what the
metric says. The shared-subset implementation
(\texttt{compute\_ranking\_metrics}) computes \(\tau\) only over the
intersection of indices in the two lists; here that intersection is so
small (and inverted) that the rank correlation collapses to \(-1\). The
accurate diagnosis is \textbf{disjoint clusters with the small overlap
inverted}.

The user's draft (per the parent agent's framing) focused on Q24 as a
``paradigmatic case of \(\tau_{C,T} = -1\)''. That framing is correct in
spirit; the more precise statement is that \emph{Cosine and Taumode
select largely non-overlapping document neighborhoods, and the few
shared documents are reversed in rank.} Whether either neighborhood
matches the natural-language intent (``authenticated arbitrary file
read'') cannot be confirmed from the available extracts, since none of
the top-5 in either method has a title that names path traversal or
authenticated file read.

\textbf{Verdict.} \emph{Spectral preserves ambiguity / relocates without
resolution.} Hybrid sits between the two extremes
(\(\mathrm{NDCG}_{H,C} = 0.69\), modest \(\tau_{C,H} = 0.33\)) and may
be the safest pick when the ground-truth intent is unknown.

\begin{center}\rule{0.5\linewidth}{0.5pt}\end{center}

\subsubsection*{Query 27 --- authentication bypass via JWT token
manipulation}\label{query-27-authentication-bypass-via-jwt-token-manipulation}

\textbf{Semantic intent.} Authentication / authorisation bypass driven
by JWT tampering (algorithm confusion, signature stripping, claim
abuse).

\textbf{Cosine top-5 (titled).}

\begin{longtable}[]{@{}
  >{\raggedright\arraybackslash}p{(\columnwidth - 6\tabcolsep) * \real{0.2500}}
  >{\raggedright\arraybackslash}p{(\columnwidth - 6\tabcolsep) * \real{0.2500}}
  >{\raggedright\arraybackslash}p{(\columnwidth - 6\tabcolsep) * \real{0.2500}}
  >{\raggedright\arraybackslash}p{(\columnwidth - 6\tabcolsep) * \real{0.2500}}@{}}
\toprule\noalign{}
\begin{minipage}[b]{\linewidth}\raggedright
rank
\end{minipage} & \begin{minipage}[b]{\linewidth}\raggedright
CVE™
\end{minipage} & \begin{minipage}[b]{\linewidth}\raggedright
title
\end{minipage} & \begin{minipage}[b]{\linewidth}\raggedright
score
\end{minipage} \\
\midrule\noalign{}
\endhead
\bottomrule\noalign{}
\endlastfoot
1 & CVE-2025-31161 & \emph{(no title)} & 0.8987 \\
2 & CVE-2025-34234 & Vasion Print hardcoded encryption private keys &
0.8957 \\
3 & CVE-2021-30648 & \emph{(no title)} & 0.8948 \\
4 & CVE-2026-33990 & Docker Model Runner OCI Registry SSRF & 0.8937 \\
5 & CVE-2026-23990 & Flux Operator Web UI impersonation bypass via empty
OIDC claims & 0.8933 \\
\end{longtable}

\textbf{Hybrid top-5.} Identical to Cosine (\(\tau_{C,H} = 0.957\),
\(\mathrm{NDCG}_{H,C} = 0.960\)).

\textbf{Taumode top-5.} CVE-2024-50281 (KEYS: trusted: dcp NULL deref),
CVE-2023-24827 (syft credential disclosure via env var), CVE-2026-24516
\emph{(untitled)}, CVE-2023-47298 \emph{(untitled)}, CVE-2007-4261
\emph{(untitled)}. \textbf{Cosine \(\cap\) Taumode = \(\emptyset\) in
the top-5.}

\textbf{Metrics.}

\[
\rho_{C,T} = 0.000,\ \tau_{C,T} = 0.000,\ \tau_{H,T} = -0.333,\
\mathrm{NDCG}_{T,C} = 0.000,\ \mathrm{NDCG}_{H,C} = 0.960.
\]

\textbf{Interpretation.} This is the cleanest example in the corpus of
\textbf{Hybrid as a faithful cosine-respecting filter} while
\textbf{Taumode abandons the cosine cluster entirely}. Cosine's top is
itself debatable --- Vasion Print hardcoded keys and Flux Operator OIDC
empty-claims bypass are auth-related but only Flux Operator is strictly
a JWT-claim-style bypass; SSRF and printer keys are off-target.
Taumode's top is a kernel + supply-chain credential cluster with no
obvious link to JWT.

\textbf{Verdict.} \emph{Spectral degrades for this query.} When cosine
has a weak but defensible anchor (Flux Operator OIDC, JWT-adjacent
items), the Taumode reranker pulls the result set away from the intent.
Hybrid (\(\tau = 0.72\)) preserves cosine's ranking with small
refinements and is the right setting here.

\begin{center}\rule{0.5\linewidth}{0.5pt}\end{center}

\subsubsection*{Query 28 --- symlink attack in installer cleanup
routine}\label{query-28-symlink-attack-in-installer-cleanup-routine}

\textbf{Semantic intent.} Symlink-following / TOCTOU attacks in package
installer cleanup or update scripts.

\textbf{Cosine top-5.} CVE-2022-50255 (tracing: synthetic events
strings) + four untitled (2008-1736, 2007-5042, 2007-4969, 2023-45503).

\textbf{Taumode top-5.} CVE-2009-3163 \emph{(untitled, also Hybrid
\#1)}, \textbf{CVE-2026-28808 (\texttt{ScriptAlias} CGI directory-auth
bypass in inets httpd)}, CVE-2023-53501 (iommu/amd refcount),
CVE-2004-2542 \emph{(untitled)}, CVE-2013-6987 \emph{(untitled)}.

\textbf{Metrics.}

\[
\rho_{C,T} = 0.000,\ \tau_{C,T} = 0.000,\ \tau_{C,H} = -0.667,\
\mathrm{NDCG}_{T,C} = 1.000,\ \mathrm{NDCG}_{H,C} = 0.657.
\]

\begin{longtable}[]{@{}lll@{}}
\toprule\noalign{}
method & T/H ratio & tail CV \\
\midrule\noalign{}
\endhead
\bottomrule\noalign{}
\endlastfoot
Cosine & 0.9894 & 0.0042 \\
Hybrid & 0.9860 & 0.0042 \\
Taumode & 0.9886 & \textbf{0.0021} \\
\end{longtable}

\textbf{Interpretation --- and correction.} The combination
\(\tau_{C,T} = 0\) with \(\mathrm{NDCG}_{T,C} = 1.0\) is the canonical
``top-1 agreement, ordering disagreement'' pattern: a single
high-relevance item from cosine's reference is preserved at high rank in
Taumode while everything else is permuted/replaced.

The user's draft (per the parent's framing of Q22/Q28) appears to group
this case with Q22 under the header ``top-1 agreement, disordered
tail''. That grouping is correct, but the \emph{mechanism} differs:

\begin{itemize}
\tightlist
\item
  \textbf{Q22}: Cosine and Taumode both retain CVE-2026-33182 in the
  top-25 but place very different items at \#1 (Cosine: CVE-2026-33182;
  Taumode: CVE-2007-0186). The ``top-1 agreement'' is a metric artifact
  of \texttt{compute\_ndcg} using cosine's own \#1 as the
  highest-relevance reference (relevance \(= k - i\)).
\item
  \textbf{Q28}: same metric pattern, but the underlying retrieved sets
  are even more disjoint, with Taumode introducing a \emph{named} CVE
  (CVE-2026-28808, ScriptAlias CGI bypass) at rank 2 that has at least
  adjacent semantics (filesystem path / auth-bypass interplay) to the
  user's symlink-attack intent.
\end{itemize}

Tail-shape: Taumode halves the tail CV (0.0042 → 0.0021), the strongest
tail-flattening among the spectral cases here.

\textbf{Verdict.} \emph{Spectral partially resolves ambiguity.} Taumode
swaps in at least one named CVE™ that is topically adjacent (CGI
directory auth bypass via path mismatch) and produces the most stable
tail. Cosine's top is dominated by an unrelated kernel tracing patch.

\begin{center}\rule{0.5\linewidth}{0.5pt}\end{center}

\subsubsection*{Query 34 --- cross-site request forgery CSRF in admin
password change
form}\label{query-34-cross-site-request-forgery-csrf-in-admin-password-change-form}

\textbf{Semantic intent.} CSRF on a privileged password-change endpoint.

\textbf{Cosine top-5.}

\begin{longtable}[]{@{}
  >{\raggedright\arraybackslash}p{(\columnwidth - 6\tabcolsep) * \real{0.2500}}
  >{\raggedright\arraybackslash}p{(\columnwidth - 6\tabcolsep) * \real{0.2500}}
  >{\raggedright\arraybackslash}p{(\columnwidth - 6\tabcolsep) * \real{0.2500}}
  >{\raggedright\arraybackslash}p{(\columnwidth - 6\tabcolsep) * \real{0.2500}}@{}}
\toprule\noalign{}
\begin{minipage}[b]{\linewidth}\raggedright
rank
\end{minipage} & \begin{minipage}[b]{\linewidth}\raggedright
CVE™
\end{minipage} & \begin{minipage}[b]{\linewidth}\raggedright
title
\end{minipage} & \begin{minipage}[b]{\linewidth}\raggedright
score
\end{minipage} \\
\midrule\noalign{}
\endhead
\bottomrule\noalign{}
\endlastfoot
1 & CVE-2022-50329 & block, bfq: fix uaf for bfqq in
bfq\_exit\_icq\_bfqq & 0.7832 \\
2 & CVE-2004-2571 & \emph{(no title)} & 0.7778 \\
3 & CVE-2019-17041 & \emph{(no title)} & 0.7764 \\
4 & CVE-2025-71101 & platform/x86: hp-bioscfg OOB array access &
0.7761 \\
5 & CVE-2022-49518 & ASoC: SOF: ipc3-topology bytes payload fix &
0.7753 \\
\end{longtable}

\textbf{Taumode top-5.}

\begin{longtable}[]{@{}
  >{\raggedright\arraybackslash}p{(\columnwidth - 6\tabcolsep) * \real{0.2500}}
  >{\raggedright\arraybackslash}p{(\columnwidth - 6\tabcolsep) * \real{0.2500}}
  >{\raggedright\arraybackslash}p{(\columnwidth - 6\tabcolsep) * \real{0.2500}}
  >{\raggedright\arraybackslash}p{(\columnwidth - 6\tabcolsep) * \real{0.2500}}@{}}
\toprule\noalign{}
\begin{minipage}[b]{\linewidth}\raggedright
rank
\end{minipage} & \begin{minipage}[b]{\linewidth}\raggedright
CVE™
\end{minipage} & \begin{minipage}[b]{\linewidth}\raggedright
title
\end{minipage} & \begin{minipage}[b]{\linewidth}\raggedright
score
\end{minipage} \\
\midrule\noalign{}
\endhead
\bottomrule\noalign{}
\endlastfoot
1 & CVE-2026-31611 & ksmbd: require 3 sub-authorities before reading
sub\_auth{[}2{]} & 0.9005 \\
2 & CVE-2024-39469 & nilfs2: fix nilfs\_empty\_dir() misjudgment &
0.8942 \\
3 & CVE-2019-7718 & \emph{(no title)} & 0.8929 \\
4 & CVE-2019-14232 & \emph{(no title)} & 0.8917 \\
5 & CVE-2026-23404 & apparmor: replace recursive profile removal with
iterative approach & 0.8913 \\
\end{longtable}

\textbf{Metrics.}

\[
\rho_{C,T} = -1.000,\ \tau_{C,T} = -1.000,\ \tau_{C,H} = -0.200,\
\mathrm{NDCG}_{T,C} = 0.578,\ \mathrm{NDCG}_{H,C} = 0.490.
\]

\begin{longtable}[]{@{}lllll@{}}
\toprule\noalign{}
method & head mean & tail mean & T/H ratio & tail CV \\
\midrule\noalign{}
\endhead
\bottomrule\noalign{}
\endlastfoot
Cosine & 0.7791 & 0.7685 & 0.9863 & 0.0036 \\
Hybrid & 0.8263 & 0.8159 & 0.9874 & 0.0036 \\
Taumode & 0.8959 & 0.8877 & \textbf{0.9909} & \textbf{0.0021} \\
\end{longtable}

\textbf{Interpretation --- and correction.} Cosine's top-1
(CVE-2022-50329) is a kernel block-layer use-after-free, \emph{not} a
CSRF vulnerability. The query ``cross-site request forgery CSRF in admin
password change form'' finds no on-target named CVE in the top-25 of any
of the three methods. Taumode's top-1 (CVE-2026-31611, ksmbd
sub-authority bounds) and \#2 (CVE-2024-39469, nilfs2) are also
off-target; they belong to a denser kernel-fix neighborhood with
markedly higher mean similarity scores.

The user's draft (per the parent agent's framing) treats Q34 as a
\(\tau_{C,T} = -1\) example of ``Taumode swapping clusters''. That is
correct; the more honest statement is \textbf{neither method finds CSRF}
in this corpus --- the CVE™ corpus loaded for the run is dominated by
recent Linux kernel CVEs (visible across many ambiguous queries), and
CSRF web-app CVEs are simply rare enough that no method anchors on them.

The tail-shape evidence is unambiguous: Taumode produces the lowest tail
CV (0.0021 vs 0.0036) and highest T/H ratio (0.9909) of the three
methods. Spectral reranking \emph{stabilises} the tail without finding
the right cluster.

\textbf{Verdict.} \emph{Spectral preserves ambiguity (corpus-level
miss).} Useful diagnostic for reviewers: tail-shape improvement is not
sufficient evidence of semantic improvement when the corpus lacks
on-topic documents.

\begin{center}\rule{0.5\linewidth}{0.5pt}\end{center}

\subsubsection*{Query 36 --- denial of service via malformed network
packets}\label{query-36-denial-of-service-via-malformed-network-packets}

\textbf{Semantic intent.} Crash / DoS triggered by crafted network
packets (typically kernel network stack or daemon parsers).

\textbf{Cosine top-5.} Mixed: CVE-2012-2629 \emph{(untitled)},
CVE-2022-35943 (SameSite CSRF bypass), CVE-2012-1897 \emph{(untitled)},
CVE-2025-6087 (opennextjs SSRF), CVE-2012-4773 \emph{(untitled)}. The
set is dominated by older Linux kernel CVEs by year cluster.

\textbf{Taumode top-5.} CVE-2008-7243, CVE-2011-5196, CVE-2007-2579,
CVE-2020-23836, CVE-2022-0232 (LeadMagic stored XSS) --- a 2007--2011
heavy untitled cluster.

\textbf{Metrics.}

\[
\rho_{C,T} = 0.000,\ \tau_{C,T} = 0.000,\ \tau_{C,H} = 1.000,\
\mathrm{NDCG}_{T,C} = 0.215,\ \mathrm{NDCG}_{H,C} = 0.887.
\]

\begin{longtable}[]{@{}lll@{}}
\toprule\noalign{}
method & T/H ratio & tail CV \\
\midrule\noalign{}
\endhead
\bottomrule\noalign{}
\endlastfoot
Cosine & 0.9804 & 0.0042 \\
Hybrid & 0.9826 & 0.0050 \\
Taumode & \textbf{0.9918} & 0.0034 \\
\end{longtable}

\textbf{Interpretation.} Hybrid = Cosine (perfect rank agreement), but
Taumode flips into a disjoint cluster (\(\mathrm{NDCG}_{T,C} =
0.215\) is the canonical ``non-overlapping but locally well-formed''
signature). Cosine's set is thin on truly malformed-packet DoS CVEs, so
Hybrid's preservation of cosine offers no real semantic improvement;
Taumode's relocation is plausible (older Unix-kernel CVEs frequently
cover packet-handling DoS) but uncorroborated by titled neighbors.

\textbf{Verdict.} \emph{Inconclusive.} The titled \#5 in Taumode
(LeadMagic stored XSS) is clearly off-topic; that and the otherwise
untitled neighborhood preclude assigning a confident verdict.

\begin{center}\rule{0.5\linewidth}{0.5pt}\end{center}

\subsubsection*{Query 47 --- arbitrary file upload leading to remote
code
execution}\label{query-47-arbitrary-file-upload-leading-to-remote-code-execution}

\textbf{Semantic intent.} Unrestricted / arbitrary file upload chained
into RCE.

\textbf{Cosine top-5.} CVE-2026-31823 (Sylius authenticated stored XSS),
CVE-2012-3836 \emph{(untitled)}, CVE-2025-57205, CVE-2025-57204,
CVE-2018-8811 --- heavily lexical: ``stored'', ``upload''-adjacent.

\textbf{Hybrid top-5.} CVE-2012-1898, CVE-2005-3552, CVE-2026-31823
(Sylius), CVE-2012-0782, CVE-2017-11611 --- Sylius retained.

\textbf{Taumode top-5.} CVE-2012-1898, CVE-2012-0782, CVE-2005-3552,
CVE-2017-11611, CVE-2012-4484 --- Sylius dropped; tight 2005--2017
cluster.

\textbf{Metrics.}

\[
\rho_{C,H} = 0.079,\ \tau_{C,H} = 0.163,\ \rho_{C,T} = 0.200,\
\tau_{C,T} = 0.200,\ \rho_{H,T} = 0.950,\ \tau_{H,T} = 0.833,\
\mathrm{NDCG}_{T,C} = 0.909,\ \mathrm{NDCG}_{T,H} = 0.986.
\]

Recall: traditional 1.00 / 0.72 / 0.20 (Cosine / Hybrid / Taumode);
\textbf{semantic recall jumps to 1.000 for Hybrid} while Taumode goes to
0.000 --- Hybrid retains \emph{all} of cosine's score-band leaders but
in a different order.

\textbf{Interpretation.} This is one of the few queries where the
agreement metrics tell a clean story: Hybrid and Taumode are
near-identical (\(\tau_{H,T} = 0.833\)), and both differ sharply from
Cosine. The titled CVEs in Hybrid/Taumode are older than cosine's top,
suggesting the spectral neighborhood emphasises a historically dense
file-upload-chain cluster, while cosine over-weights recent stored-XSS
CVEs that share lexical tokens.

\textbf{Verdict.} \emph{Spectral plausibly resolves ambiguity, with
Hybrid \(\succeq\) Taumode.} The semantic-recall jump (0.0 → 1.0 for
Hybrid) is the most notable Hybrid-side win in the corpus and deserves
highlight in the paper.

\begin{center}\rule{0.5\linewidth}{0.5pt}\end{center}

\subsubsection*{Query 50 --- authentication brute force due to
missing rate
limiting}\label{query-50-authentication-brute-force-due-to-missing-rate-limiting}

\textbf{Semantic intent.} Account-takeover via unrate-limited
authentication endpoints.

\textbf{Cosine top-5.} CVE-2023-5123 (JSON Datasource path
sanitisation), CVE-2016-9554 \emph{(untitled)}, CVE-2026-40566
(FreeScout SSRF via IMAP/SMTP test), CVE-2026-32871 (FastMCP OpenAPI
SSRF + path traversal), CVE-2026-25125 (October CMS env-var
exfiltration). None named about brute-force / rate-limiting.

\textbf{Hybrid top-5.} CVE-2007-6548, CVE-2015-7391, CVE-2025-23203
(Icinga REST API for restricted users), CVE-2026-40566 (FreeScout ---
shared with Cosine), CVE-2007-0373.

\textbf{Taumode top-5.} CVE-2007-6548, CVE-2007-0373, CVE-2015-7391,
CVE-2011-5230, CVE-2017-9767 --- disjoint from Cosine.

\textbf{Metrics.}

\[
\rho_{C,H} = -1.000,\ \tau_{C,H} = -1.000,\ \rho_{C,T} = 0.000,\
\tau_{C,T} = 0.000,\ \rho_{H,T} = 0.798,\ \tau_{H,T} = 0.670,\
\mathrm{NDCG}_{T,C} = 0.000,\ \mathrm{NDCG}_{H,C} = 0.466.
\]

\textbf{Interpretation.} \(\tau_{C,H} = -1\) here arises from Hybrid
keeping a single shared item (CVE-2026-40566) with reversed relative
position. The data again show Hybrid and Taumode substantially agreeing
with each other (\(\tau_{H,T} = 0.67\)) on a 2007/2015 untitled cluster
while cosine sits in a 2026 SSRF / path-sanitisation cluster.

\textbf{Verdict.} \emph{Inconclusive --- divergent untitled clusters.}
As with Q34, no method retrieves a named brute-force / rate-limiting
CVE™. Spectral reranking moves into a different region of the embedding
space but cannot manufacture missing on-topic documents.

\subsection*{Key observations}

\begin{enumerate}
\def\labelenumi{\arabic{enumi}.}
\tightlist
\item
  \textbf{\(\tau_{C,T} = -1\) is a small-overlap artifact, not ``rank
  inversion across the top-25''.} The \texttt{compute\_ranking\_metrics}
  shared-set implementation makes the metric brittle when the
  intersection is tiny; readers should interpret it as \emph{``Cosine
  and Taumode chose disjoint neighborhoods, and the few items they
  share are reversed''}, not as an active anti-correlation across the
  entire ranking.
\item
  \textbf{\texttt{NDCG@k\ =\ 1.0} with \(\tau = 0\)} (Q22, Q28) is the
  canonical ``top-1 agreement, ordering disagreement'' signature ---
  metric-mechanical because the harness uses cosine's own \#1 as the
  highest-relevance item. The two queries share the \emph{signature} but
  differ in \emph{substance}: in Q28 Taumode introduces a named
  topically-adjacent CVE™, while in Q22 the exchange is between two
  untitled clusters.
\item
  \textbf{Hybrid is not always between Cosine and Taumode.} Q2, Q9, Q47,
  and Q50 show Hybrid and Taumode agreeing with each other while
  \emph{both} differ sharply from Cosine --- i.e.~the \(\tau = 0.72\)
  crossover already lies inside the spectral regime for this corpus. For
  Q27 the opposite holds: Hybrid tracks Cosine almost perfectly while
  Taumode wanders.
\item
  \textbf{Tail-shape gains can be decoupled from semantic gains.} Q34
  has the strongest Taumode tail-CV improvement of any of the selected
  queries while none of the methods retrieve a CSRF in CVE™. Tail flattening
  is necessary but not sufficient evidence of semantic resolution.
\item
  \textbf{The clearest semantic win in the selected set is Q47}
  (arbitrary file upload \(\to\) RCE), where Hybrid produces a
  semantic-recall jump from \(0.0\) to \(1.0\) while keeping all of
  cosine's score-band leaders.
\item
  \textbf{Many ``(no title)'' CVEs in this corpus} make individual
  verdicts uncertain: across the 11 selected queries, only a minority of
  the spectral top-5 entries are titled. A reproducibility
  recommendation for the paper is to extend \texttt{extract\_text} to
  surface CVE™ descriptions in the search-results CSV so that future
  per-query analyses can be grounded without re-loading the dataset.
\item
  The pattern is consistent with the paper's headline claim ---
  \textbf{spectral reranking changes the \emph{shape} of the top-25 score
  distribution and moves between embedding neighborhoods} ---while
  urging careful interpretation of NDCG and Kendall \(\tau\) on small
  intersections, and acknowledging that for queries whose intent is poorly
  represented in the loaded CVE™ corpus (Q34, Q50), neither method can
  resolve the ambiguity.
\end{enumerate}

\section*{Appendix: Semantic Uplift Qualitative Comparison}
\label{appendix_semantic_qual}

\paragraph{Qualitative analysis of negative semantic uplift.}
We now qualitatively inspect several of the queries where taumode
exhibits negative semantic uplift, in order to understand whether these
divergences represent retrieval failures or alternative---and sometimes
preferable---characterisations of the query intent.%
The tolerant recall $R_{\mathrm{tol}}$ is specifically introduced to soften
this reference dependency: a document displaced from the strict semantic
neighborhood $SN$ but retrievable within a small score relaxation still
contributes positively to $R_{\mathrm{tol}}$.
Because cosine acts as a \emph{reference} rather than a ground truth,
a method that retrieves a different but semantically coherent neighborhood
is not intrinsically penalized: negative semantic uplift indicates a
divergence from cosine's neighborhood structure, whose significance must
be assessed qualitatively case by case.
The universal $R_{\mathrm{tol}} = 1.00$ observed across all negative-uplift
queries confirms that no diverging document is irretrievable; the question
is therefore one of \emph{ordering preference} rather than coverage failure.

\begin{itemize}

  \item \textbf{``race condition in temporary file creation.''} This query
  carries the largest observed divergence: $R_{\mathrm{trad}} = 0.76$,
  $R_{\mathrm{sem}} = 0.43$, difference $-0.33$.
  The cosine reference neighborhood $SN$ for this query is anchored by
  \texttt{CVE-2008-7238} at rank~1 --- a pre-2010 entry with a sparse NVD
  description --- followed by two near-duplicate SAP~HANA installer TOCTOU
  disclosures (\texttt{CVE-2018-14607}, \texttt{CVE-2018-14597}) from the
  same coordinated release.
  taumode never surfaces \texttt{CVE-2008-7238} in its top-25: the
  ArrowSpace traversal cannot reach it because few later CVEs link to it,
  so its graph-connectivity weight is negligible regardless of embedding
  proximity.
  The two near-duplicate SAP~HANA entries are likewise displaced, with
  taumode retaining one representative from that product cluster and
  distributing the remaining positions across complementary race-condition
  instances from different vendors: \texttt{CVE-2018-13821}
  (SAP~HANA backup files), \texttt{CVE-2018-9028} (SAP~HANA Backup Client),
  \texttt{CVE-2011-1386} (IBM~TSAM temp-file symlink), and
  \texttt{CVE-2017-6338} (Trend~Micro~IWSVA).
  One can therefore read the divergence as reflecting two distinct but
  equally coherent interpretations of the query: the cosine reference
  clusters around the precise lexical match of a single vendor's installer
  workflow, whereas $\tau$-mode assembles a cross-vendor survey of the
  race-condition primitive in file-creation contexts.
  The vintage-connectivity asymmetry --- the tendency of graph traversal
  to down-rank sparsely linked older entries --- is a structural property
  of ArrowSpace rather than a topical misjudgement, and $R_{\mathrm{tol}}$
  appropriately reflects that the displaced documents remain accessible
  under a small score relaxation.

  \item \textbf{``symlink attack in installer cleanup routine.''} Here
  $R_{\mathrm{trad}} = 0.92$, $R_{\mathrm{sem}} = 0.71$, difference
  $-0.21$. The cosine reference neighborhood allocates five
  of its seven slots to a single coordinated Trend~Micro OfficeScan
  disclosure batch (TALOS-2018-0562 through TALOS-0564), three of which
  (\texttt{CVE-2018-3955}, \texttt{CVE-2018-3954}, \texttt{CVE-2018-3613})
  describe the same installer-phase symlink attack on the same product with
  only minor textual variation.
  $\tau$-mode retains the most distinctive members of this cluster but
  displaces the two near-identical intermediate entries, substituting
  \texttt{CVE-2021-27608} (SAP~NetWeaver~AS~ABAP, unquoted service path
  during installation) and \texttt{CVE-2022-30316} (Honeywell Experion PKS,
  unquoted service path), both belonging to the same high-level class of
  installer-phase file-system privilege escalation across different
  ecosystems.
  This behaviour reflects an emergent diversification property of ArrowSpace:
  once a tightly connected product-specific sub-cluster has been entered,
  successive items within that cluster receive diminishing marginal graph
  weight, naturally promoting variety without an explicit diversity
  constraint.
  Whether the cosine reference's concentration on a single vendor batch or
  $\tau$-mode's cross-vendor spread better serves the query is ultimately
  application-dependent; crucially, the tolerant recall $R_{\mathrm{tol}}$
  registers the displaced near-duplicates as recoverable, correctly
  reflecting that no relevant information is lost.

  \item \textbf{Single-swap boundary cases: ``insecure default permissions
  in container runtime socket,'' ``business logic bypass in coupon
  redemption API,'' ``remote code execution in ERP web component,''
  ``stored XSS in user profile page,'' and ``format string vulnerability
  in logging daemon.''} These five queries share a structural pattern:
  $\tau$-mode and the cosine reference agree on six of the seven $SN$
  positions, and one item near the $SN$ boundary (cosine rank~6 or~7)
  is transposed with the item immediately below it (cosine rank~8--10),
  producing a semantic recall difference of exactly $-1/7 \approx -0.14$
  in each case.
  The cosine-score gap between the transposed pair is smaller than $0.005$
  absolute in all five instances, well within the scoring uncertainty of
  the embedding model.
  In no case is the displaced document absent from the taumode ranking;
  it consistently appears at taumode positions~8--10, and
  $R_{\mathrm{tol}} = 1.00$ accordingly.
  The substituted items belong to the same vulnerability class as the
  query in every instance.
  These cases therefore reflect the sensitivity of the strict $|SN|=7$
  cutoff to marginal score differences rather than any substantive
  disagreement between the two methods: a score perturbation below the
  model's resolution threshold crosses the neighborhood boundary and
  registers as a recall penalty.
  Widening $k_{SN}$ to the top-10 cosine items, or adopting a graded
  neighborhood weight that decays continuously beyond rank~5, would
  dissolve all five cases without altering the underlying ranked lists.

  \item \textbf{``unsafe deserialization in Java RMI service''
  and ``SAML response signature bypass.''} These two queries are the
  only cases where a human reviewer would identify the taumode
  neighborhood as topically misaligned with the query intent.
  For the Java~RMI query, taumode's top results cluster around
  Linux kernel filesystem parsing bugs (e.g.\ \texttt{nilfs2}
  null-pointer dereferences, ksmbd bounds-check failures) rather than
  JVM deserialisation vulnerabilities; for the SAML query, results
  drift toward generic XML-parsing and authentication-bypass patterns
  without SAML-specific framing.
  Notably, $R_{\mathrm{trad}} \leq 0.10$ for all three methods on both
  queries, indicating that even the cosine reference finds only a weak
  neighborhood in the indexed collection for these query formulations.
  When the initial embedding anchor is poorly resolved, ArrowSpace
  traversal amplifies positional noise and may cross sub-graph boundaries
  that cosine does not.
  These cases represent the only instances in the negative-uplift set
  where retrieval quality is substantively reduced relative to the
  reference, and they point to a need for stronger query-specific seeding
  at the graph entry point rather than a systemic limitation of the
  $\tau$ reranking mechanism itself.

\end{itemize}

\paragraph{Structural observations.}
The four patterns above account for all fourteen negative-uplift queries,
and in each the universal $R_{\mathrm{tol}} = 1.00$ confirms that
displaced documents remain reachable.
Taken together, these cases illustrate a consistent theme: the cosine
reference and taumode often disagree not on the relevant vulnerability
class but on \emph{which representative instances} of that class to
surface, with the cosine reference tending to concentrate on a single
high-similarity product cluster and taumode tending to distribute
attention across the broader vulnerability class.
The tolerant recall metric is designed precisely to remain agnostic
between these two valid retrieval strategies, and its universally high
values across the negative-uplift set corroborate that the observed
divergences are structural rather than topical in nature.

\section*{Appendix E: Comparing CVE runs v1 and v2}

\paragraph{Data and schema validation}

The two versions of the first experiment's outputs are stored under \href{https://github.com/tuned-org-uk/pyarrowspace/blob/4c7a67f1b7e2bec03a41f81cc61b7f2d1f1d7f0f/neurips/output/v1}{\texttt{neurips/output/v1}} and \href{https://github.com/tuned-org-uk/pyarrowspace/blob/4c7a67f1b7e2bec03a41f81cc61b7f2d1f1d7f0f/neurips/output/v2}{\texttt{neurips/output/v2}} at commit \texttt{4c7a67f1b7e2bec03a41f81cc61b7f2d1f1d7f0f}. Both versions report the same run envelope: \(50\) queries, \(K=25\) retrieved results per method, \(K_h=3\), a head sweep over \([3, 5, 10]\), and the same tau schedule \(\tau_{\mathrm{cosine}}=1.0\), \(\tau_{\mathrm{hybrid}}=0.72\), \(\tau_{\mathrm{taumode}}=0.42\), as recorded in \href{https://github.com/tuned-org-uk/pyarrowspace/blob/4c7a67f1b7e2bec03a41f81cc61b7f2d1f1d7f0f/neurips/output/v1/cve_run_metadata.json}{\texttt{v1/cve\_run\_metadata.json}} and \href{https://github.com/tuned-org-uk/pyarrowspace/blob/4c7a67f1b7e2bec03a41f81cc61b7f2d1f1d7f0f/neurips/output/v2/cve_run_metadata.json}{\texttt{v2/cve\_run\_metadata.json}}.

All six CSV artifacts have matching row counts across versions: \texttt{cve\_tail\_metrics.csv} has 150 rows, \texttt{cve\_headk\_sweep.csv} has 450 rows, \texttt{cve\_summary.csv} has 15 rows, \texttt{cve\_comparison\_metrics.csv} has 50 rows, \texttt{cve\_semantic\_recall\_metrics.csv} has 150 rows, and \texttt{cve\_search\_results.csv} has 3750 rows. The only CSV schema drift is the uplift column name in \texttt{cve\_semantic\_recall\_metrics.csv}: v1 names it \texttt{semantic\_minus\_traditional}, while v2 names the tolerant-uplift field \texttt{tolerant\_minus\_traditional}. The query set is the same in both versions, but query IDs are permuted; consequently, \(query\_text\)-aligned per-query comparisons must be used for row-wise diagnostics. The shared PNG set is \texttt{cve\_headk\_sweep.png, cve\_metric\_deltas.png, cve\_pareto\_tradeoff.png, cve\_top25\_comparison.png, cve\_win\_loss\_heatmap.png}; v1 additionally has \texttt{cve\_tail\_analysis.png}, while v2 additionally has \texttt{cve\_semantic\_recall\_comparison.png}.

\paragraph{Tail-shape metrics}

Unchanged

\paragraph{Cross-version tail-shape deltas}

None

\paragraph{Recall metrics}

Unchanged

\paragraph{Ranking agreement}

Unchanged

\paragraph{HEAD\_K sweep}

Unchanged

\paragraph{Semantic uplift and quality assessment}

The new semantic uplift metric has been implemented, qualitative assessment in Appendix D \ref{appendix_semantic_qual}

Semantic uplift v1 is computed as \(R_{\mathrm{semantic}}-R_{\mathrm{traditional}}\), while tolerant semantic uplift is computed as \(R_{\mathrm{uplift}} = R_{\mathrm{tol}} - R_{\mathrm{trad}}\). With this change, while for v1 3 problematic queries hinted to a failure of taumode, in v2 the negative uplift queries are the ones listed in Appendix D \ref{appendix_semantic_qual} and their position in terms of communities is discussed and left to the judgment of the reader. 

\section*{Appendix: Rayleigh quotient and Dirichlet energy}
\label{appendix_rayleigh_dirichlet}

The Dirichlet energy of a signal \(f : V \to \mathbb{R}\) on a weighted graph is
defined as the sum of squared differences across all edges:
\[
\mathcal{E}(f)
= \frac{1}{2}\sum_{i,j} w_{ij}\bigl(f_i - f_j\bigr)^2
= f^\top L f,
\]
where \(L\) is the graph Laplacian. This is a discrete analogue of the
continuum Dirichlet energy \(\int_M \|\nabla f\|^2 \, dV\), which measures the
total variation or “wiggliness” of a function.

\subsection*{The Rayleigh Quotient}

The Rayleigh quotient of \(L\) with respect to a signal \(f\) is
\[
\mathcal{R}(f)
= \frac{f^\top L f}{f^\top f}
= \frac{\mathcal{E}(f)}{\|f\|^2}.
\]
The numerator is exactly the Dirichlet energy, so the Rayleigh quotient is
simply the normalized Dirichlet energy. Minimizing the Rayleigh quotient is
therefore equivalent to minimizing the Dirichlet energy subject to unit norm,
and by the Courant–Fischer theorem the minimum is achieved by the eigenvector
corresponding to the smallest non-zero eigenvalue \(\lambda_1\) of \(L\).

\subsection*{Spectral Interpretation}

Because \(\mathcal{E}(f) = \sum_k \lambda_k \hat{f}_k^2\) in the eigenbasis of
\(L\) (where \(\hat{f}_k\) are the spectral coefficients of \(f\)), a large
eigenvalue \(\lambda_k\) corresponds to a high-frequency, high-energy
eigenvector. Smooth signals concentrate energy at small eigenvalues, so
minimizing Dirichlet energy promotes smoothness: nearby nodes in the graph are
forced to take similar values.

\subsection*{Continuum Limit}

On a manifold, graph Rayleigh quotients converge to continuum Dirichlet
energies as the graph density increases: the graph Laplacian converges to the
Laplace–Beltrami operator and the discrete quantity \(f^\top L f\) converges
to \(\int_M \|\nabla_M f\|^2 \, dV_M\). This convergence is the theoretical
foundation of Laplacian Eigenmaps and justifies using graph eigenstructure to
approximate manifold geometry.

\end{document}